\documentclass[a4paper,12pt,twocolumn]{article}

\usepackage{array}
\usepackage{amsmath}
\usepackage{amssymb}
\usepackage{amstext}
\usepackage{amsfonts}
\usepackage{amsbsy}
\usepackage{amsopn}
\usepackage{amsthm}
\usepackage{amscd}
\usepackage{amsxtra}
\usepackage[comma,authoryear]{natbib}
\usepackage[dvips]{graphicx,color,epsfig}
  \newcommand{\R}{{\sf R\hspace*{-0.9ex}%
    \rule{0.15ex}{1.5ex}\hspace*{0.9ex}}}

\usepackage{url}

\begin{document}
\bibliographystyle{abbrvnat}
\title{High-dimensional switches and the modeling of cellular differentiation}
\date{}
\author{  Olivier Cinquin $^{1,2,3,*}$ \& Jacques Demongeot $^{3}$\\ {\small 1: CoMPLEX, UCL} \\ {\small 2:Department of Anatomy and Developmental Biology}\\ {\small University College London} \\ {\small Gower Street, London WC1E 6BT, U.K.} \\ {\small 3:Laboratoire TIMC-IMAG-CNRS UMR 5525, Universit\'e Joseph Fourier, Grenoble 1}\\ {\small Facult\'e de M\'edecine, Domaine de la Merci, 38706 La Tronche, France} \\ * Correspondence: o.cinquin@ucl.ac.uk \\ Journal of Theoretical Biology (in press, 2005)}
\maketitle

\section{Abstract}
Many genes have been identified as driving cellular differentiation, but because of their complex interactions, the understanding of their collective behaviour requires mathematical modelling. Intriguingly, it has been observed in numerous developmental contexts, and particularly hematopoiesis, that genes regulating differentiation are initially co-expressed in progenitors despite their antagonism, before one is upregulated and others downregulated. 

We characterise conditions under which 3 classes of generic "master regulatory networks", modelled at the molecular level after experimentally-observed interactions (including bHLH protein dimerisation), and including an arbitrary number of antagonistic components, can behave as a "multi-switch", directing differentiation in an all-or-none fashion to a specific cell-type chosen among more than 2 possible outcomes. bHLH dimerisation networks can readily display coexistence of many antagonistic factors when competition is low (a simple characterisation is derived). Decision-making can be forced by a transient increase in competition, which could correspond to some unexplained experimental observations related to Id proteins; the speed of response varies with the initial conditions the network is subjected to, which could explain some aspects of cell behaviour upon reprogramming.

The coexistence of antagonistic factors at low levels, early in the differentiation process or in pluripotent stem cells, could be an intrinsic property of the interaction between those factors, not requiring a specific regulatory system.

Abbreviations: bHLH, basic Helix-Loop-Helix, Id, Inhibitor of Differentiation

Keywords: multistationarity, cellular differentiation, cellular reprogramming, bHLH dimerization

\section{Introduction}

It has long been recognised that cellular differentiation could result from epigenetic memory, controlled  by the dynamical properties of the same system, present with an identical structure in all cells \citep{delbruck}, rather than from a progressive, irreversible loss of differentiation potential; a fundamental property of such a control system would be the presence of positive feedback circuits \citep{conjecture_thomas,plahte_proof,snoussi_proof,gouze_proof,CinquinO:02,soule}. Indeed, pioneer experiments showed that the genomes of some differentiated cell types retain the capacity to regenerate a whole organism \citep{gurdon_cloning1,gurdon_cloning2}, and more recent experiments have strengthened the view that there is extensive plasticity in cell-fate determination (reviewed by \citealp{BlauHM:91}, \citealp{BlauHM:99}, and \citealp{TheiseND:02}).

Bistable switches have been given a thorough theoretical investigation \citep{CherryJL:00}, and have been constructed \emph{de novo} \citep{GardnerTS:00} or modified \citep{OzbudakEM:04}. There is evidence, discussed in section \ref{biological_aspects}, that cells undergoing differentiation sometimes face commitment decisions which involve more than two possible outcomes, but switches involving more than two variables have not been given extensive attention (we are not aware of any generic mathematical model that addresses cellular differentiation, with more than two possible outcomes). In the following, we discuss the relevance of these high-dimensional switches to the modeling of cellular differentiation, and investigate the properties of different molecular models, evaluating them with the current knowledge of the mechanisms of cellular differentiation. In particular, we test whether these models are able to display a coexistence of antagonistic factors at low levels, as decision-making with increased expression levels could be a relevant model of differentiation.

\subsection{Biological aspects}
\label{biological_aspects}

\subsubsection{Some commitments are irreducible to binary steps}
\label{decision_cascades}

Cellular differentiation is often envisioned as a temporal cascade of decisions, by which cells restrict their potential fate further and further, until they reach a unique fate. It has been argued that each of these decisions is binary \citep{BrownG:88,SternbergPW:89,KalettaT:97,LinR:98}. However, recent studies of hematopoeisis strongly suggest otherwise \citep{RothenbergEV:99}, and point to models in which many cross-antagonising factors compete with each other (see below), receiving activation or inhibition from extracellular signals, leading to the progressive up-regulation of one specific factor, and down-regulation of all others. The hypothesis that decisions are more complex than binary is also supported by the fact that the same cell type can be obtained by different developmental pathways \citep{RothenbergEV:99}. 

Apart from hematopoiesis, two systems have been described which seem to clearly involve a 3-outcome decision, irreducible to a sequence of 2 binary decisions: cells in the \emph{C. elegans} hemaphrodite germline are directed to mitosis, differentiation as sperm, or differentiation as oocyte \citep{EllisRE:95}, and founder cells of Drosophila mesoderm are directed to specific dorsal muscle or pericardial cell phenotypes by 3 mutually-repressive genes \citep{JaglaT:02}.

Finally, in at least two instances of neural development, fate choices between a great diversity of possible outcomes have been shown, and are unlikely to be mediated by a series of binary commitments. This is the case of olfactory development \citep{SerizawaS:00,EbrahimiFA:00}, which does not involve genetic rearrangements \citep{EgganK:04}, and of the regulation of hundreds of alternatively-spliced transcripts of a single gene in the Drosophila brain \citep{NevesG:04}.

\begin{figure}
\begin{center}
\includegraphics[width=3in]{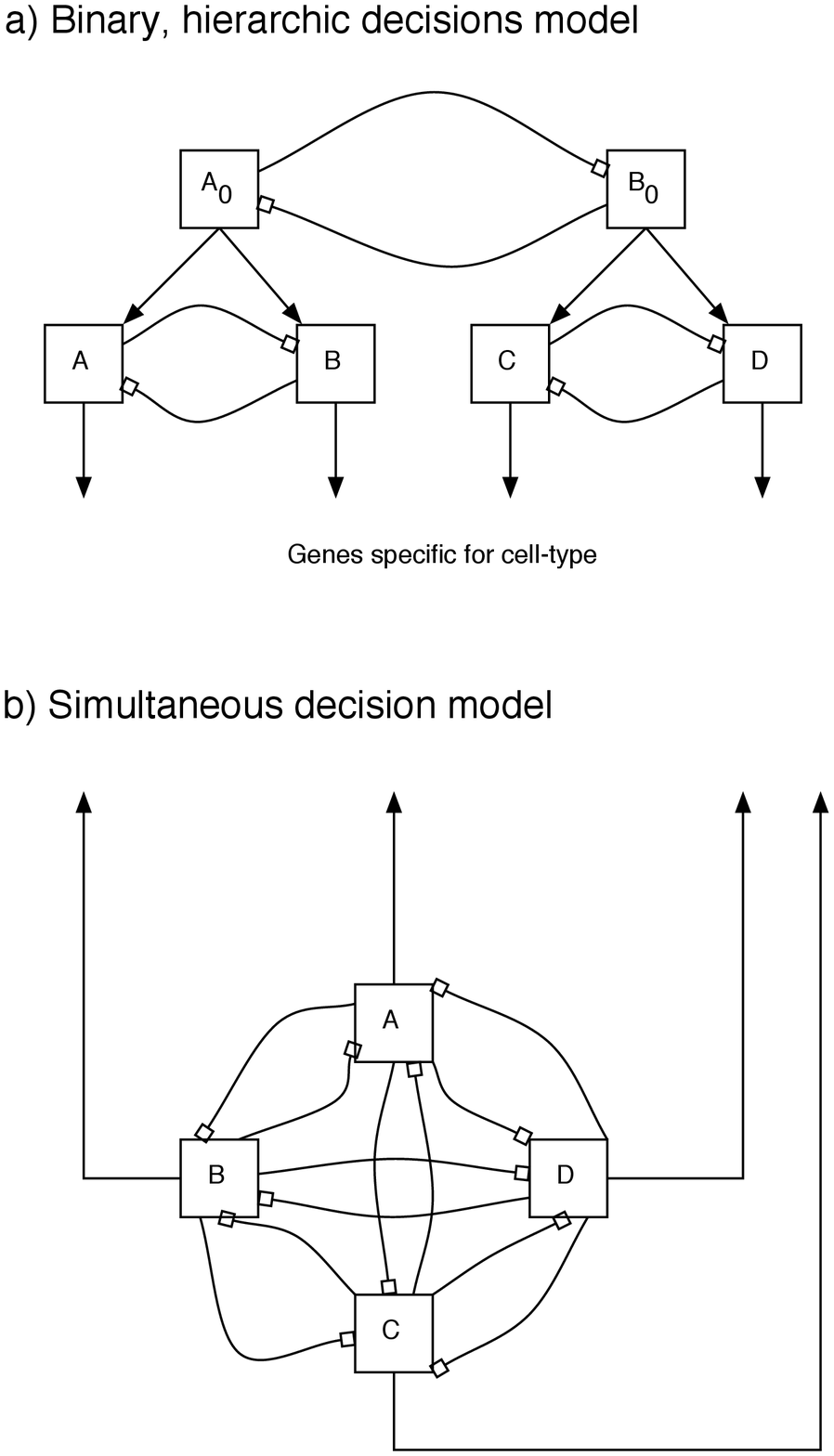}
\caption{Arrows represent activation, and squares inhibition. Adapted from \citet{CinquinO:02}.}
\label{diff_models}
\end{center}
\end{figure}

Thus, it appears that model a, depicted in Figure \ref{diff_models}, is not the only possibility, and that model b of Figure \ref{diff_models} should also be taken into account.

Having shown that high-dimensional switches are necessary for the mathematical modeling of some developmental decisions, we now turn to the way their structure should be modeled: differentiation factors are often antagonistic (section \ref{antagonism_section}), which doesn't prevent them from being sometimes coexpressed (section \ref{coexpression_section}), and modulation of the interaction strength is a way differentiation is regulated (section \ref{differentiation_regulation}). The basis for a mathematical formulation of the models is provided in section \ref{math_models}.

\subsubsection{Antagonism between differentiation factors}
\label{antagonism_section}

Antagonism between genes driving differentiation to different fates has been repeatedly established; often, enforced expression of a differentiated phenotype, whether by specific misexpression of a gene, or fusion of cells with different phenotypes, also leads to repression of the previous phenotype (repression of alternative fates has been proposed to be an essential mechanism of differentiation, reviewed by \citealp{CoryS:99}). The idea of competition is reinforced by dose-dependency effects, shown for example by comparison of heterozygous and homozygous mutants, heterokaryon studies, or knock-down mutations (\citealp{WeintraubH:93}, \citealp{McDevittMA:97}, reviewed by \citealp{OrkinSH:00}; \citealp{CrittendenSL:02}), by monoallelic expression of a gene such as Pax5 \citep{NuttSL:99}, and by dosage effects of interacting bHLH proteins \citep{ZhuangY:96}. These effects argue that boolean models, in which a specific master gene would be turned on, initiate transcription of cell-type specific genes, and repress all other fates, are insufficient.

Competition between cell-fate determining factors has also been documented at the molecular level, for example in the case of neurogenesis, where bHLH proteins play a major role in specifying neural subtypes \citep{ChienCT:96,BrunetJF:99}. \citet{GowanK:01} have identified a network of 3 cross-repressive bHLH proteins (although not all possible cross-repressions have been characterised).
 \citet{BriscoeJ:00} have also shown that a cross-repressive gene network reads out the Shh gradient in the neural tube. Two sets of two cross-repressing genes have been identified, with a possibility that there is a larger, totally cross-repressive network (all the possible interactions do not seem to have been assessed yet). 
 The competition can also happen by physical interaction between the factors, rather than by cross-repression of transcription: in hematopoeisis, GATA-1, which drives erythroid and megakaryocytic differentiation  \citep{KulessaH:95,VisvaderJE:92,IwasakiH:03}, and PU-1, a transcription factor essential for the expression of myleoid-specific genes (reviewed by \citealp{ZhangDE:96}), as well as B-cell specific genes \citep{ChenH:96}, suppress each other's activity by physical interaction \citep{RekhtmanN:99,ZhangP:99,NerlovC:00}. This seems to be a general phenomenon in hematopoiesis (\citealp{HuM:97}, reviewed by \citealp{CrossMA:97,EnverT:98}).
 
 In addition to repressing other genes, cell-fate determining factors often enhance their own expression; it has been proposed that this is a common property of "master switches" \citep{RothenbergEV:99}. 

\subsubsection{Coexpression of antagonistic factors}
\label{coexpression_section}

Coexpression of antagonistic genes has been shown both for closely-related lineages (for example coexpression of antagonistic hematopoeisis-related genes, \citealp{MiyamotoT:02}, \citealp{AkashiK:03}, \citealp{YeM:03}, reviewed by \citealp{OrkinSH:03}, transient prespore expression of a prestalk-specific gene in Dictyostelium, \citealp{JermynK:95}, coexpression of lineage-specific genes in pancreas development, \citealp{ChiangMK:03}, and coexpression of neurogenic genes, \citealp{RalluM:02,PieraniA:01,BriscoeJ:00}, although the latter may be due to a transient effect of the misexpression method), and between more distantly-related lineages (for example expression of neural markers by hematopoietic precursors, \citealp{GoolsbyJ:03}). 

A  semi-quantitative analysis of the expression of many hematopoietic genes was performed by \citet{AkashiK:00}, showing that lineage-specific (and antagonistic) genes were co-expressed at low levels in precursors, before respective upregulation and downregulation (see \citealp{RothenbergEV:00}, \citealp{ZhuJ:02}, for reviews). At an earlier stage of development, markers for different germ layers are also transiently co-expressed \citep{WardleFC:04}.

\subsubsection{Regulation of differentiation}
\label{differentiation_regulation}

Some proteins have been shown to have regulative effects on differentiation in many different cellular contexts, and would thus prove interesting to incorporate in models of cellular differentiation. 
\begin{itemize}

\item{Id proteins, ubiquitously expressed during development, seem to act as inhibitors of cell differentiation, by sequestering ubiquitously-expressed class A bHLH proteins, preventing class B bHLH to form A-B heterodimers, which are transcriptionally active (\citealp{BenezraR:90,GarrellJ:90,EllisHM:90}, reviewed by \citealp{NortonJD:98,NortonJD:00}), and by preventing DNA binding \citep{O'ToolePJ:03}; see \citet{MassariME:00} for a precise classification of HLH proteins. Twist can act in the same way \citep{SpicerDB:96}, or in another, more direct way, by binding to class MyoD \citep{HamamoriY:97}.}

\item{Hes1, a bHLH protein, seems in many cases to be essential in the maintenance of an undifferentiated state \citep{KageyamaR:00}; its effect can be mediated either by active repression, which involves the recruitment of Groucho, or by passive repression, which involves hetero-dimerisation with other bHLH proteins. }

\item{The PUF family of proteins represses the expression of many genes by regulating their mRNA stability \citep{WickensM:02}, and has been proposed to have the ancestral function of maintaining proliferation of stem cells; in \emph{C. elegans}, sex-determination genes are regulated by PUF members.}

\item{NF-$\kappa$B has been shown to inhibit differentiation of mesenchymal cells, by destabilisation of the transcripts of Sox9 and MyoD, two transcription factors involved in different differentiation pathways \citep{SitcheranR:03}.}

\end{itemize}

All these differentiation-inhibiting proteins have a negative effect on the strength of transcription of genes which are essential in cell-fate determination. The models presented below suggest that modulation of the transcription strength of proteins involved in cell-fate determination could allow for an initial co-existence of many antagonistic factors, followed by up-regulation of one factor at the expense of others, as the transcription strength is increased.

\subsection{Mathematical models}
\label{math_models}

The models studied here have an arbitrary number of components. Each variable represents the intracellular concentration of a differentiation factor (called "switch element" in the following), which enhances its own expression and represses that of all others (the system is symmetrical, in that any element has the same relationship with all others, and in that all elements share a common set of parameters). The models can represent different forms of biological interactions. The terminology used below is that of transcriptional control: each factor is supposed to be a protein, which enhances the transcription of its own mRNA, and represses the transcription of the mRNAs for other switch elements, with or without physical interaction with other factors; as a simplification, the translation step is not taken into account in the model, and proteins are thus supposed to act directly on each other's concentrations. There is evidence that translational regulation can play a major role in some cases \citep{translational_control,OkanoH:02}. In the following models, different forms of post-transcriptional control (by means of regulation of mRNA stability, or translation of the proteins), can be represented in the same way as transcriptional control. Downregulation of cytokine receptors has been observed prior to commitment \citep{KondoM:00}, and downregulation of receptors promoting expression of competing factors could also be accounted for by the following models.

3 kinds of models are studied below: 
\begin{itemize}
\item{Mutual inhibition with autocatalysis: all switch elements repress one another, and enhance their own expression. This is one of the simplest models one can think of that is able to achieve dominant expression of each of its elements, depending on the initial conditions.}
\item{Mutual inhibition with autocatalysis, \emph{and leak}: the same as the previous, with a supplementary term that represents an identical, basal level of expression, which is independent of any element of the network. This could correspond for example to a gene upstream in the differentiation hierarchy, which "primes" the lower level of the differentiation network, as has been proposed within the hematopoietic differentiation network (\citealp{YeM:03}, reviewed by \citealp{OrkinSH:03}).}
\item{bHLH dimerisation: based on the class A/class B bHLH dimerisation discussed above.}
\end{itemize}

The first two models can be viewed as a generic representation of the interactions between switch elements, while the third is based on an explicit assumption. All are formulated according to standard kinetics.

These models are cell-autonomous, and do not take into account "differentiation cues" that cells receive. The models could be extended to take into account either different initial conditions, leading to various differentiated states, or different biases of the network (for example by providing a higher basal expression level of one of the factors).

\section{Results}

\subsection{Mutual inhibition with autocatalysis}

Each switch element is supposed to undergo non-regulated degradation (modeled as exponential decay, with an arbitrary speed 1), and transcription/translation with a relative speed $\sigma$. Each element positively auto-regulates itself, and represses expression of others, with a cooperativity $c$. Calling $x_i$ the concentration of each switch element, the corresponding equations are, for $n$ elements:

{\large
\begin{align}
\nonumber
\frac{\mathrm{d}x_1}{\mathrm{d}t} & =-x_1+\frac{\sigma x_{1}^{c}}{1+\Sigma_{i=1}^{n} x_{i}^\mathrm{c}}\\
\label{equations_autocat}
& \dots \\
\nonumber
\frac{\mathrm{d}x_n}{\mathrm{d}t} & =-x_n+\frac{\sigma x_{n}^{c}}{1+\Sigma_{i=1}^{n} x_{i}^\mathrm{c}} 
\end{align}
}

The analysis is restricted to $c\ge1$, as there is only one steady state (0) if $c<1$. The results presented in appendix \ref{mia} show that it is possible for one switch element to be on and others off (for $\sigma>2$), but that if there is some cooperativity in the system (\emph{ie} $c>1$), it is impossible for more than 1 element to be on at the same time. On the contrary, if there is no cooperativity ($c=1$), simulations show that a multitude of equilibria exist and are stable (switch elements in the "on" state can even coexist at different concentrations). Thus, the multistability behaviour of this system is tunable only by changes in the strength of the cooperativity, a mechanism which seems to be of unlikely biological relevance.

\subsection{Mutual inhibition with autocatalysis, and leak}

The model is the same as previously, except that each element has a "leaky" expression, modelled as a constant production term $\alpha$. The equations become:

{\large
\begin{align}
\nonumber
\frac{\mathrm{d}x_1}{\mathrm{d}t} & =- x_1+\frac{\sigma x_{1}^{c}}{1+\Sigma_{i=1}^{n} x_{i}^\mathrm{c}} + \alpha\\
& \dots \\
\nonumber
\frac{\mathrm{d}x_n}{\mathrm{d}t} & =- x_n+\frac{\sigma x_{n}^{c}}{1+\Sigma_{i=1}^{n}x_{i}^\mathrm{c}} + \alpha
\end{align}
}

The system is interesting only for $c>1$ (see appendix \ref{mial}). If the leak is small, it doesn't have a major effect on the system, except when the cooperativity is close to 1: as shown in appendix \ref{mial}, it is impossible for more than one switch element to be "on", at a much higher level than the leak level $\alpha$.

\begin{figure}
\begin{center}
\includegraphics[width=2.5in]{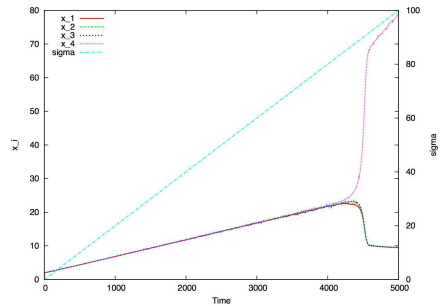}
\end{center}
\caption{Time evolution of the concentrations of 4 switch elements ($x_1$ to $x_4$), for the model with mutual inhibition with autocatalysis, and leak, with the transcription strength $\sigma$ being gradually increased over time. The 4 elements are initially coexpressed at an identical level, which increases with $\sigma$; when $\sigma$	 reaches a threshold level, one element is upregulated, and others are downregulated. Parameters in the simulation were $\alpha=2$ and $c=1.1$ Low, random noise was added to allow the system to escape the equilibrium as it became unstable.}
\label{increasing_sigma}
\end{figure}

Even when the cooperativity is close to 1, it is still the case that only one variable at the same time can dominate all others; in order for that to happen, the transcription strength must be sufficiently high. A simulation was performed for a cooperativity of $1.1$, with increasing $\sigma$ (see Figure \ref{increasing_sigma}). All switch elements are initially coexpressed, and once $\sigma$ becomes sufficiently high, one switch element is upregulated, and others downregulated. 

\begin{figure}
\begin{center}
\includegraphics[width=2.7in]{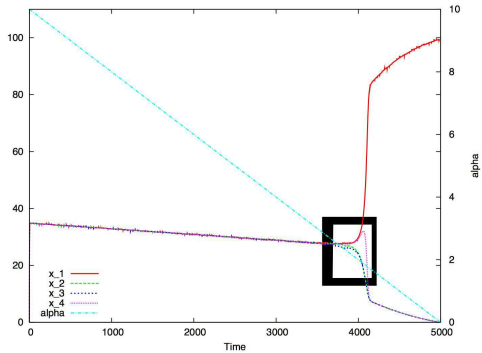}
\end{center}
\caption{Time evolution of the concentrations of 4 switch elements ($x_1$ to $x_4$), for the model with mutual inhibition with autocatalysis, and leak, with the leak level $\alpha$ being gradually decreased over time. The 4 elements are initially coexpressed at identical levels (higher than the leak $\alpha$ because of autocatalysis); when the leak reaches a threshold level, one element is upregulated, and others are downregulated. Note that the scales for the $x_i$ and for $\alpha$ are different by a factor of 11, equal to $c/(c-1)$ in this simulation. Thus, it is impossible for the curve of more than one $x_i$ to be above that of $\alpha$ \emph{at equilibrium}. Thus, in the boxed region, the system is in the process of responding to the drop in $\alpha$, and not at equilibrium. Parameters in the simulation were $\sigma=100$ and $c=1.1$ Low, random noise was added to allow the system to escape the equilibrium as it became unstable.}
\label{decreasing_leak}
\end{figure}

The same pattern of coexpression followed by exclusive expression can be achieved with a decreasing leak (see Figure \ref{decreasing_leak}), with the difference that the level of initial coexpression decreases slightly with time (this level is lower than the relative maximum transcription strength $\sigma$, but higher than the leak $\alpha$). Once the leak has become sufficiently small, exclusive upregulation occurs.

We show in appendix \ref{mial} that our models with mutual inhibition and autocatalysis, with or without leak, always converge to an equilibrium (and thus never oscillate).

\subsubsection{Effect of a perturbation}

If two switch elements are given initial values close to the resting value one would have on its own, there is a transient drop in both values, until the higher one picks up and extinguishes the other (see Figure \ref{b7}). The initial drop is less pronounced than for the bHLH dimerisation model (see below).

\begin{figure}
\begin{center}
\includegraphics[width=3.2in,angle=-90]{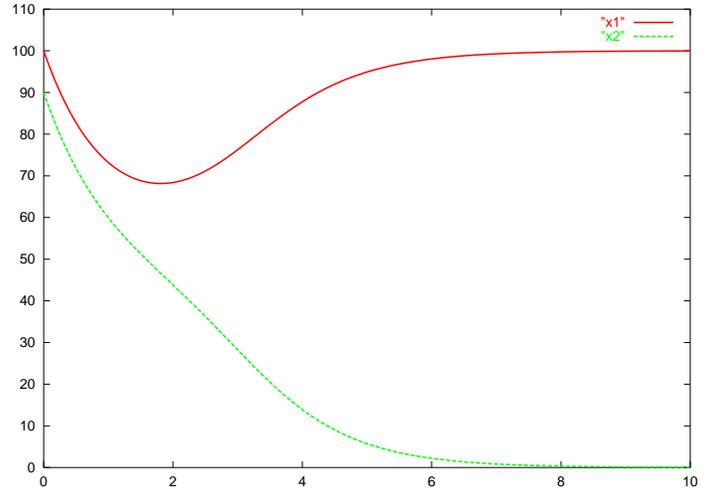}
\end{center}
\caption{Time evolution of the concentrations of two switch elements ($x_1$ and $x_2$), for the model with mutual inhibition with autocatalysis, and leak. The resting concentration when one element is on and the other off is roughly 100. Initial concentrations differ by 10. Parameters are $\sigma=100$,  $\alpha=0.02$, and $c=2$. The trajectory is essentially the same for all $\alpha<10$, and very similar for initial concentrations differing by only 1.}
\label{b7}
\end{figure}

\subsection{A model for bHLH proteins}
 
Each switch bHLH protein is supposed to bind to a common activator according to the law of mass action, with a binding constant $K_2$, and a total quantity of activator $a_t$. In turn, the heterodimer activates transcription of the corresponding switch protein only, with Hill kinetics and cooperativity 2 (as with cooperativity 1, no interesting equilibria exist, as shown in appendix \ref{bHLH_analysis_details}). The equations are:

{\Large
\begin{align}
\nonumber
\frac{\mathrm{d}x_1}{\mathrm{d}t} & =-x_1+\sigma \frac{\left( \frac{a_t x_1}{1+\Sigma_{i=1}^{n}x_i} \right)^2}  {K_2 + \left( \frac{a_t x_1}{1+\Sigma_{i=1}^{n}x_i} \right)^2} \\
& \dots \\
\nonumber
\frac{\mathrm{d}x_n}{\mathrm{d}t} & =-x_n+\sigma \frac{\left( \frac{a_t x_n}{1+\Sigma_{i=1}^{n}x_i} \right)^2}  {K_2 + \left( \frac{a_t x_n}{1+\Sigma_{i=1}^{n}x_i} \right)^2}
\end{align}
}

These equations simplify to:

$$
\frac{\mathrm{d}x_i}{\mathrm{d}t}=-x_i+\sigma \frac{x_i^2}{\alpha D^2 + x_i^2},
$$

with $D=1+\Sigma_{i=1}^nx_i$, $\sigma,\ \alpha=K_{2}/a_{t}^2 \ \in \R_{*}^{+}$

\begin{figure}
\begin{center}
\includegraphics[width=3in,angle=-90]{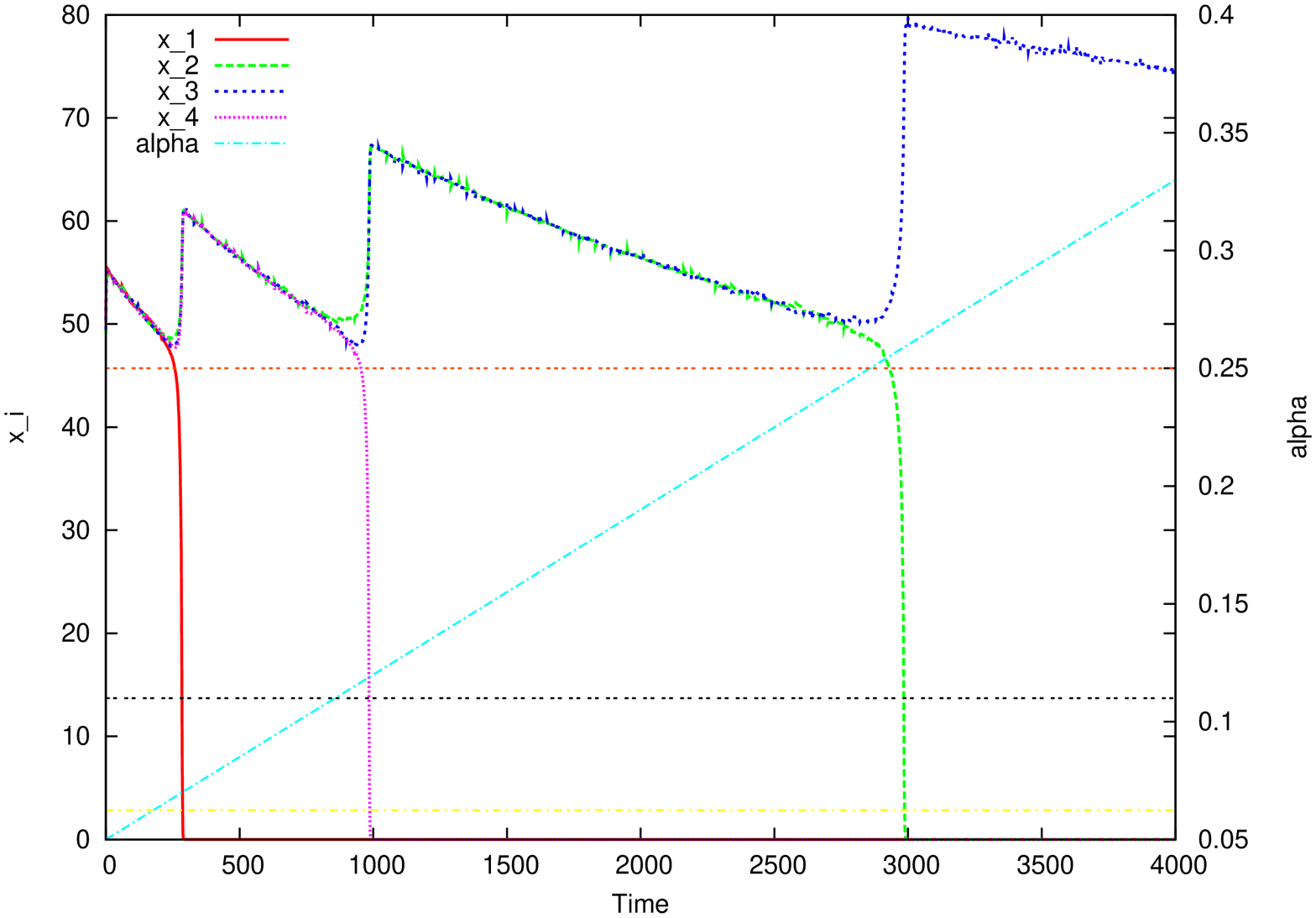}
\end{center}
\caption{Time evolution of the concentrations of 4 switch elements ($x_1$ to $x_4$), in the bHLH dimerisation model, with competition parameter $\alpha$ being gradually increased over time. The horizontal lines mark the values $\alpha=1/4^2$, $\alpha=1/3^2$, and $\alpha=1/2^2$. Each time $\alpha$ reaches one of those threshold values, one of the switch elements is repressed. Low, random noise was added to allow the system to escape equilibria as they became unstable. In this simulation $\sigma=100$.}
\label{increasing_alpha}
\end{figure}

\begin{figure}
\begin{center}
\includegraphics[width=3in]{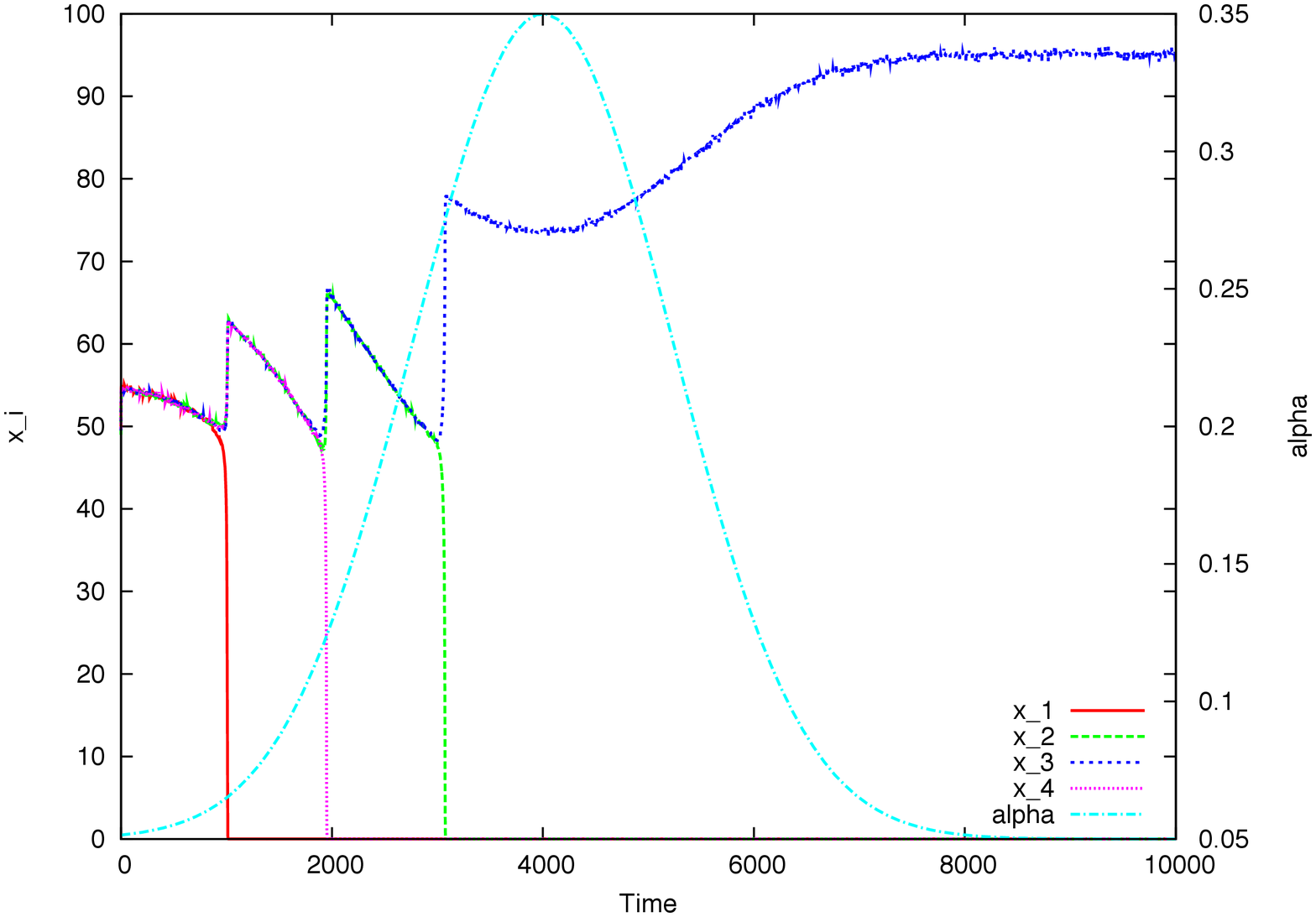}
\end{center}
\caption{Same as Figure \ref{increasing_alpha}, but with a pulse of the competition parameter $\alpha$. The initial conditions are such that the switch elements coexist for low $\alpha$; once $\alpha$ has sufficiently increased, only 1 switch element stays on, and remains on with all others off, even when $\alpha$ is brought back to its initial, low value.}
\label{bell_alpha}
\end{figure}

Parameter $\alpha$ is a measure of the harshness of the competition between switch elements (it increases if $K_2$ increases, \emph{ie} if binding to the common activator occurs at higher concentrations, and if $a_t$ diminishes, \emph{ie} if the quantity of common activator goes down). The value of $\alpha$ is essential in determining the behaviour of the system. As shown in appendix \ref{bHLH_analysis_details}, and summarised in section \ref{alpha_summary}, switch elements can coexist only if $\alpha$ is sufficiently low, \emph{ie} if the competition is not too harsh (the lower $\alpha$, the more switch elements can be non-0 at equilibrium). Figure \ref{increasing_alpha} shows a simulation with $\alpha$ being increased over time; switch elements are sharply turned off as $\alpha$ reaches threshold values. Figure \ref{bell_alpha} shows how an increase in $\alpha$ leaves only 1 switch element on, which remains exclusively on even if the competition level is relaxed to its original value.

We prove in the appendix that the system always converges to an equilibrium for $\alpha \ge 1/2$; extensive simulations have also shown this to be the case for $\alpha<1/2$.

\subsubsection{Basins of attraction and times of response}

\begin{figure}
\begin{center}
\includegraphics[width=3in]{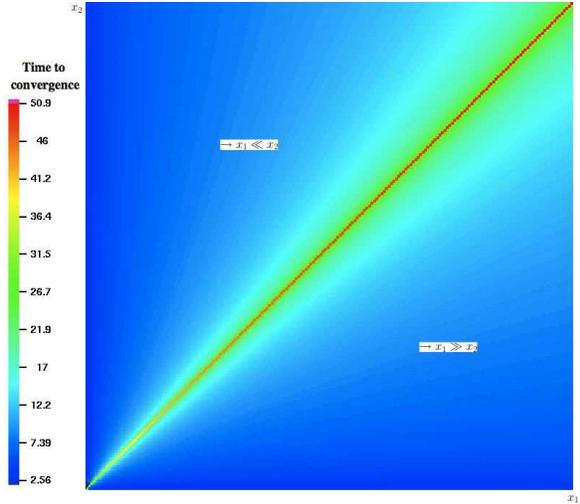}
\end{center}
\caption{Colour-coded time of convergence (as defined in Appendix \ref{convergence_times}), as a function of the initial conditions in $x_1$ and $x_2$. From the initial conditions to the left of the red region, the system converges to $x_2$ on and $x_1$ off, and the opposite for the initial conditions to the right of the red region. Parameters are $\alpha=0.4$ and $\sigma=100$. $x_1$ and $x_2$ range from 0 to 300.}
\label{alpha04}
\end{figure}

\begin{figure}
\begin{center}
\includegraphics[width=3in]{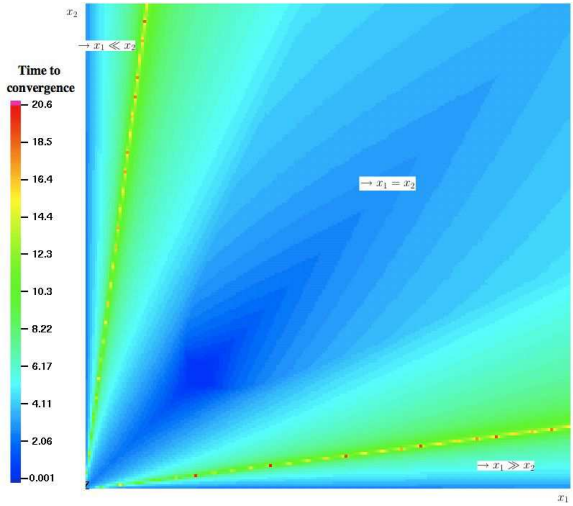}
\end{center}
\caption{Same as Figure \ref{alpha04}, but with a lower value of $\alpha$, giving a large domain of convergence to an equilibrium where $x_1$ and $x_2$ coexist. Domains of convergence are indicated, and are separated by the two yellow lines. Parameters are $\alpha=0.1$ and $\sigma=100$. $x_1$ and $x_2$ range from 0 to 300.}
\label{alpha02}
\end{figure}

\begin{figure}
\begin{center}
\includegraphics[width=3in]{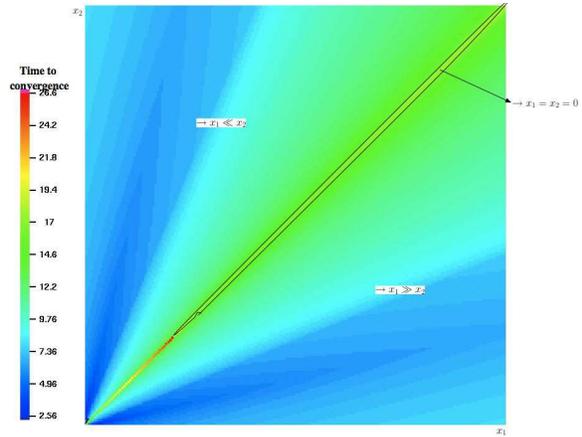}
\end{center}
\caption{Same as Figure \ref{alpha04}, but with a markedly higher value of $\alpha$. There are two main domains of convergence (to one switch element on and the other off), and a third domain of convergence to 0 (complete extinction of the switch), in a region very close to the upper part of the diagonal (for clarity reasons, the region is indicated as larger than it actually is). Parameters are $\alpha=15$ and $\sigma=100$. $x_1$ and $x_2$ range from 0 to 300.}
\label{alpha15}
\end{figure}

\begin{figure}
\begin{center}
\includegraphics[width=3in]{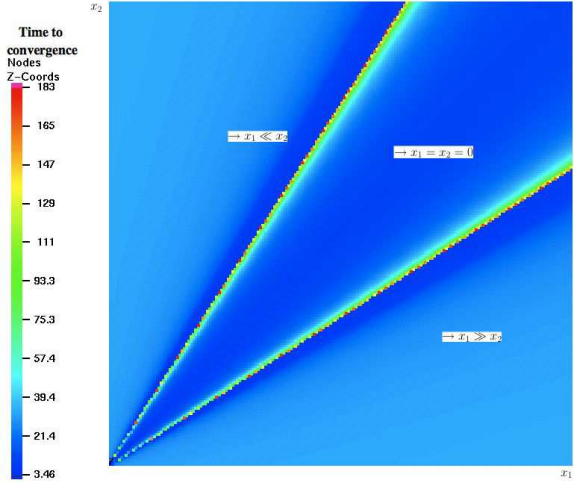}
\end{center}
\caption{Same as Figure \ref{alpha04}, with $\alpha$ close to the threshold above which $0$ is the only equilibrium. The region from which the system converges to $0$ has expanded. Parameters are $\alpha=24.75$ and $\sigma=100$. $x_1$ and $x_2$ range from 0 to 300.}
\label{alpha24}
\end{figure}

\begin{figure}
\begin{center}
\includegraphics[width=3in]{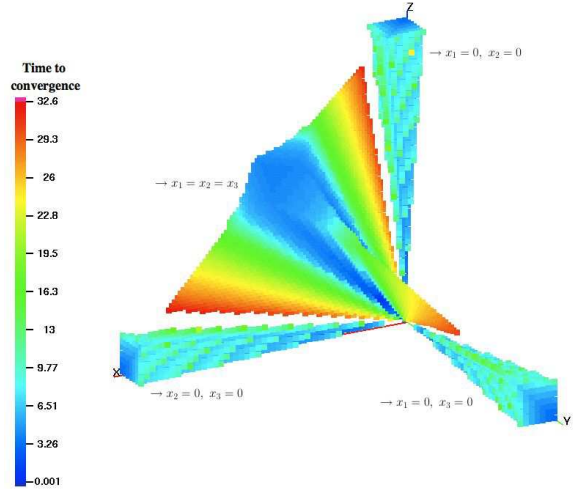}
\end{center}
\caption{Times of convergence as a function of the initial condition, for a 3-dimensional switch. 4 unconnected domains of convergence to the same equilibrium are shown. For visibility, the 3 other domains are shown in Figure \ref{3d_x5}. Parameters are $\alpha=0.1$ and $\sigma=25$. A rotation movie is available at http://www-timc.imag.fr/Olivier.Cinquin/High-dimensional\_switches\_and\_the\_modeling\_of\_cellular\_differentiation/rotating\_graphs.html }
\label{3d_x1_x2_x3_x4}
\end{figure}

\begin{figure}
\begin{center}
\includegraphics[width=3in]{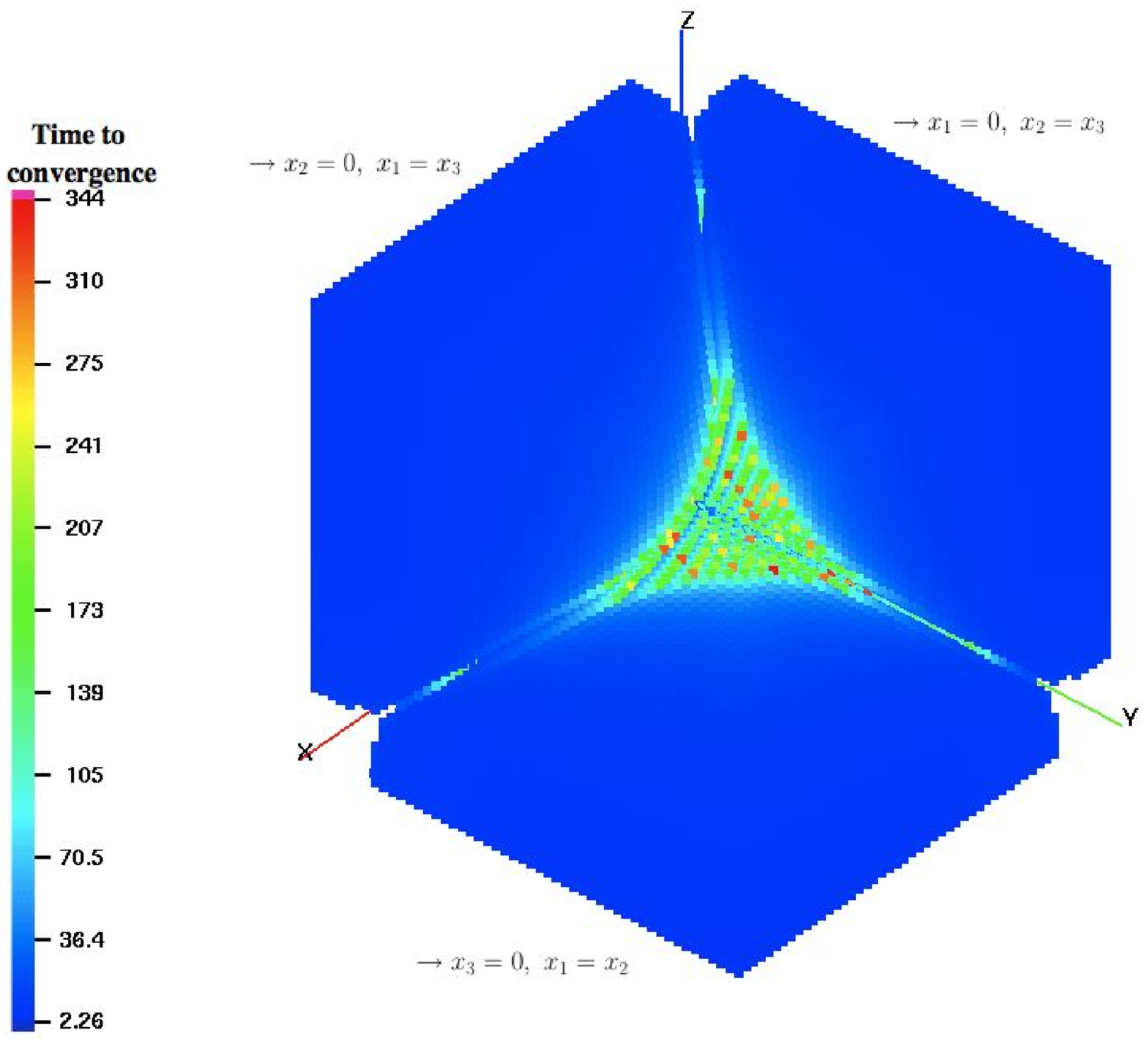}
\end{center}
\caption{Domains in which the same switch as in Figure \ref{3d_x1_x2_x3_x4} converges to a state with 2 switch elements on. A rotation movie is available at http://www-timc.imag.fr/Olivier.Cinquin/High-dimensional\_switches\_and\_the\_modeling\_of\_cellular\_differentiation/rotating\_graphs.html }
\label{3d_x5}
\end{figure}

The cell fusion and reprogramming experiments, described below in section \ref{dynamics}, would lead to a situation where a switch element, previously repressed, is  brought to a level comparable to that of another switch element which was already expressed. This corresponds to an initial situation in which two elements are not at their resting value, which could also describe the situation in cells in the process of differentiating.
For the models studied here, if 2 switch elements are competing, 3 outcomes are possible: the switch element at the higher concentration completely represses the other, both coexist and reach a non-zero equilibrium at the same value (only an element which started at the higher concentration can end up predominating), or both go to 0 (extinction). Figures \ref{alpha04} to \ref{alpha24} show which equilibrium the system converges to, as a function of the initial state, for different values of the competition parameter $\alpha$ (each domain from which the system converges to the same equilibrium is a "basin of attraction"). If there are 3 switch elements competing, there are more possibilities, as 2 or 3 elements can coexist in the "on" state. Figures \ref{3d_x1_x2_x3_x4} and \ref{3d_x5} show the basins of attraction of such a switch (the attraction basins belong to the same system, but were split on two figures to prevent the outer ones from obscuring the inner ones).

The speed at which the competition between the switch elements is carried out could be crucial in determining the dynamical properties of differentiation. We thus computed the time it takes for the system to approach its equilibrium, starting from various initial concentrations of the switch elements (that time is colour-coded in Figures \ref{alpha04} to \ref{3d_x5}). This time becomes dramatically higher when the initial conditions come close to the borders of the basins of attraction (\emph{ie} when concentrations are near a threshold around which the system converges to two or more different outcomes). The effect becomes even more pronounced when 3, rather than 2, switch elements are competing (Figures \ref{3d_x1_x2_x3_x4} and \ref{3d_x5}).

\begin{figure}
\begin{center}
$\begin{array}{c}
 \includegraphics[width=2in,angle=-90]{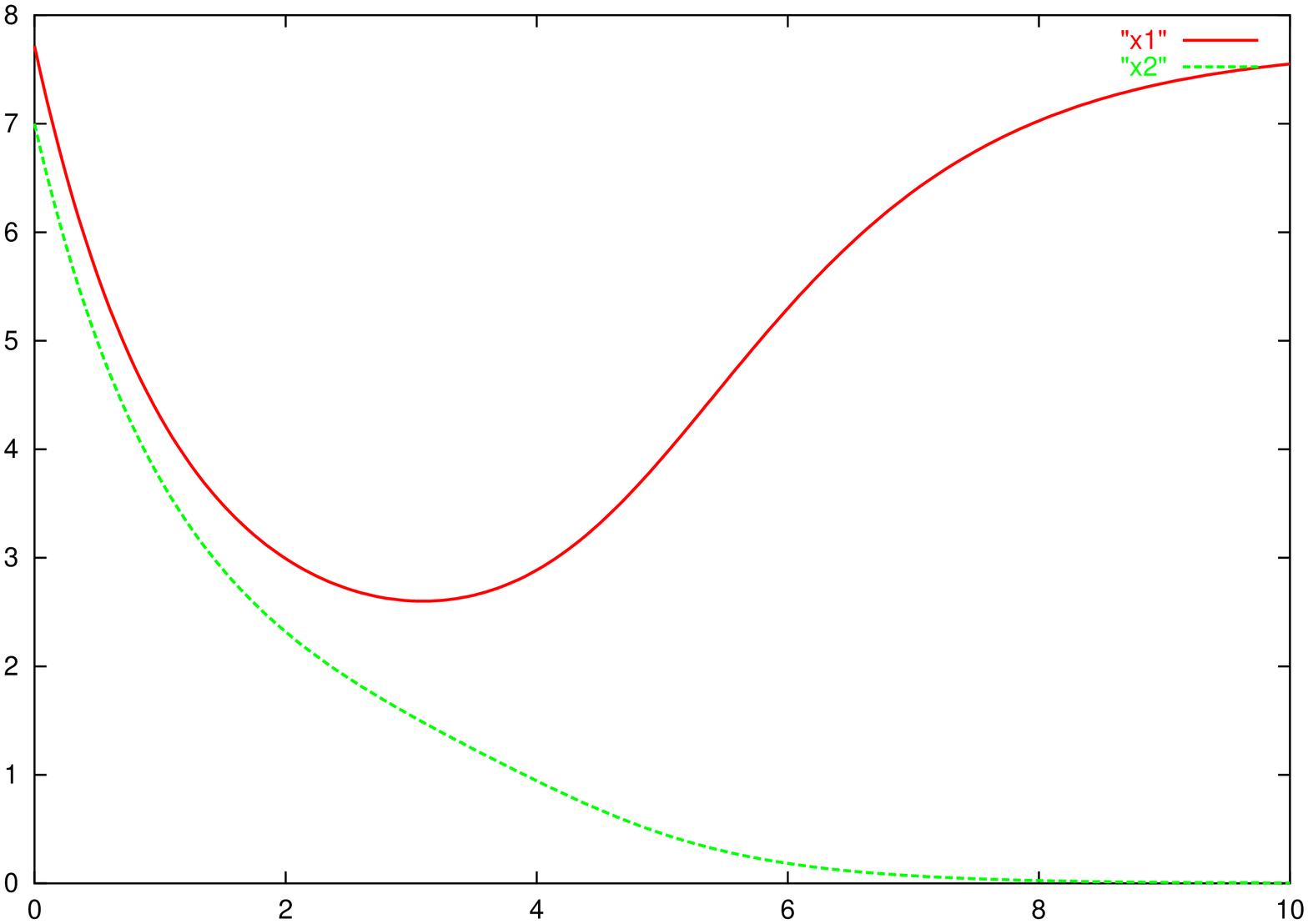} \\
\mathbf{a} \\
  \includegraphics[width=2in,angle=-90]{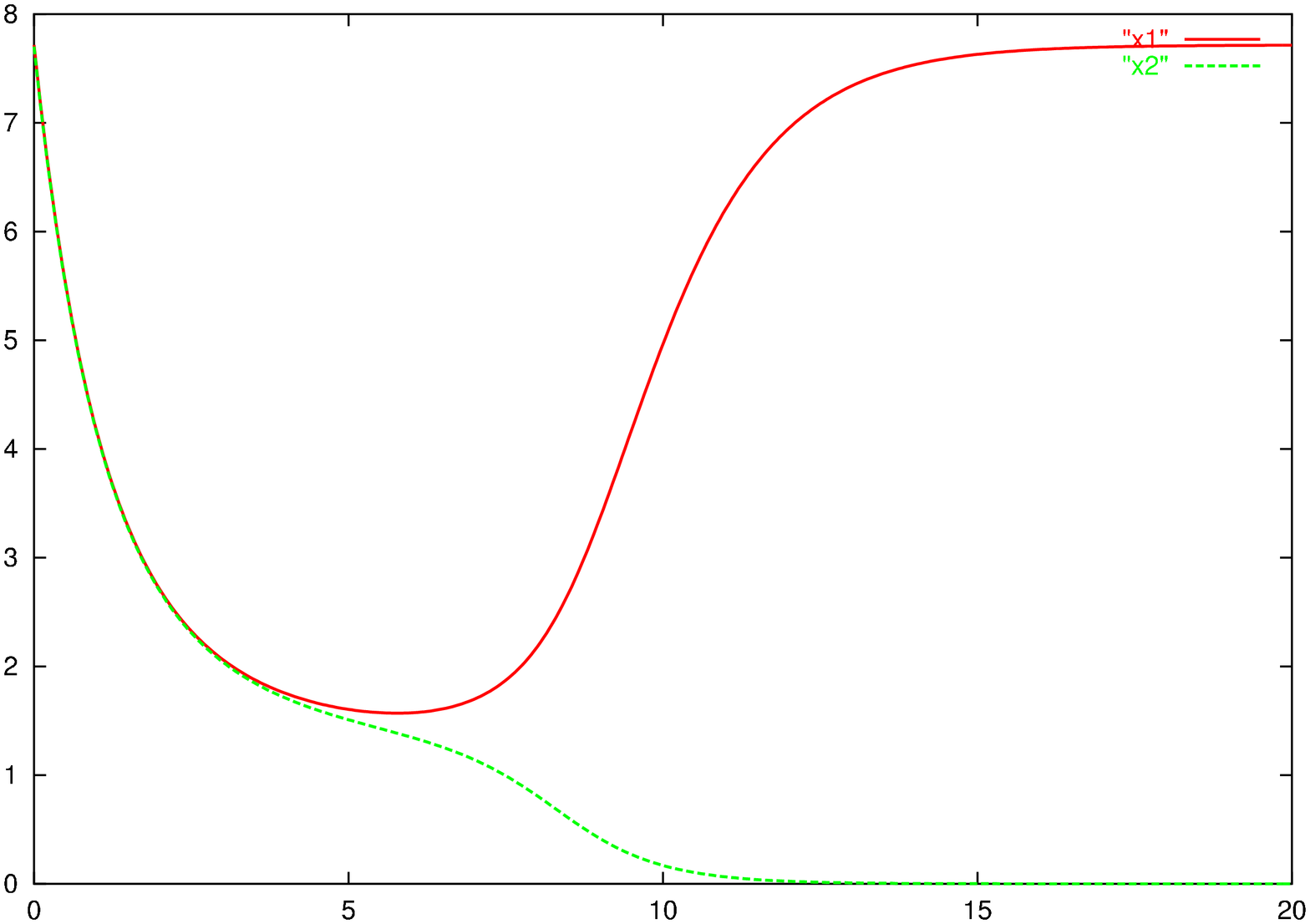} \\
\mathbf{b}
\end{array}$
\caption{Time evolution of the concentrations of two switch elements ($x_1$ and $x_2$), for the bHLH dimerisation model. The resting concentration when one element is on and the other off is roughly 8. Initial concentrations differ by 0.7 ($\mathbf{a}$), or 0.1 ($\mathbf{b}$). Notice the differences in scales. Parameters are $\alpha=50$ and $\sigma=500$.}
\label{ic_resting}
\end{center}
\end{figure}

\begin{figure}
\begin{center}
$\begin{array}{c}
 \includegraphics[width=2in,angle=-90]{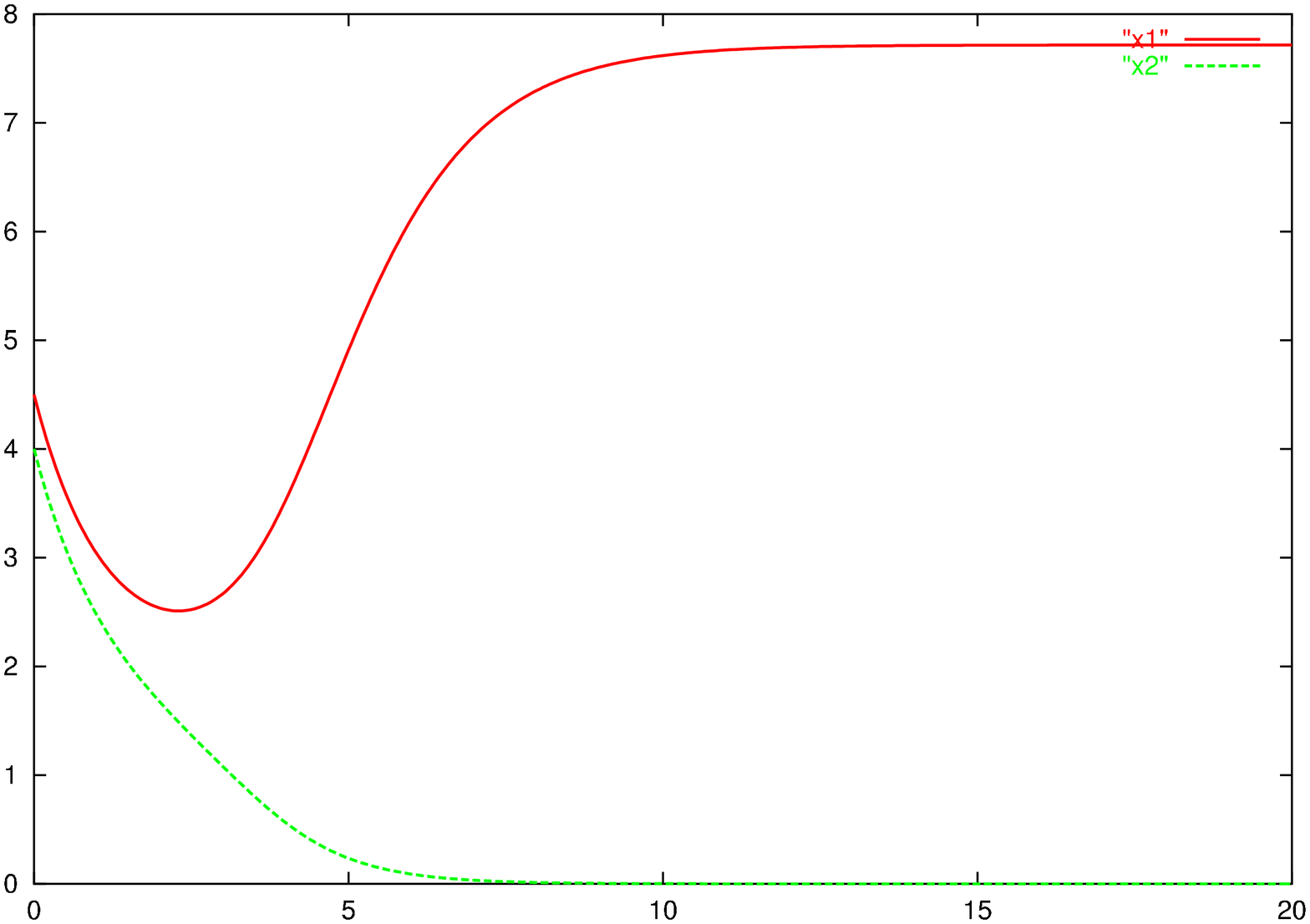} \\
 \mathbf{a} \\
  \includegraphics[width=3in]{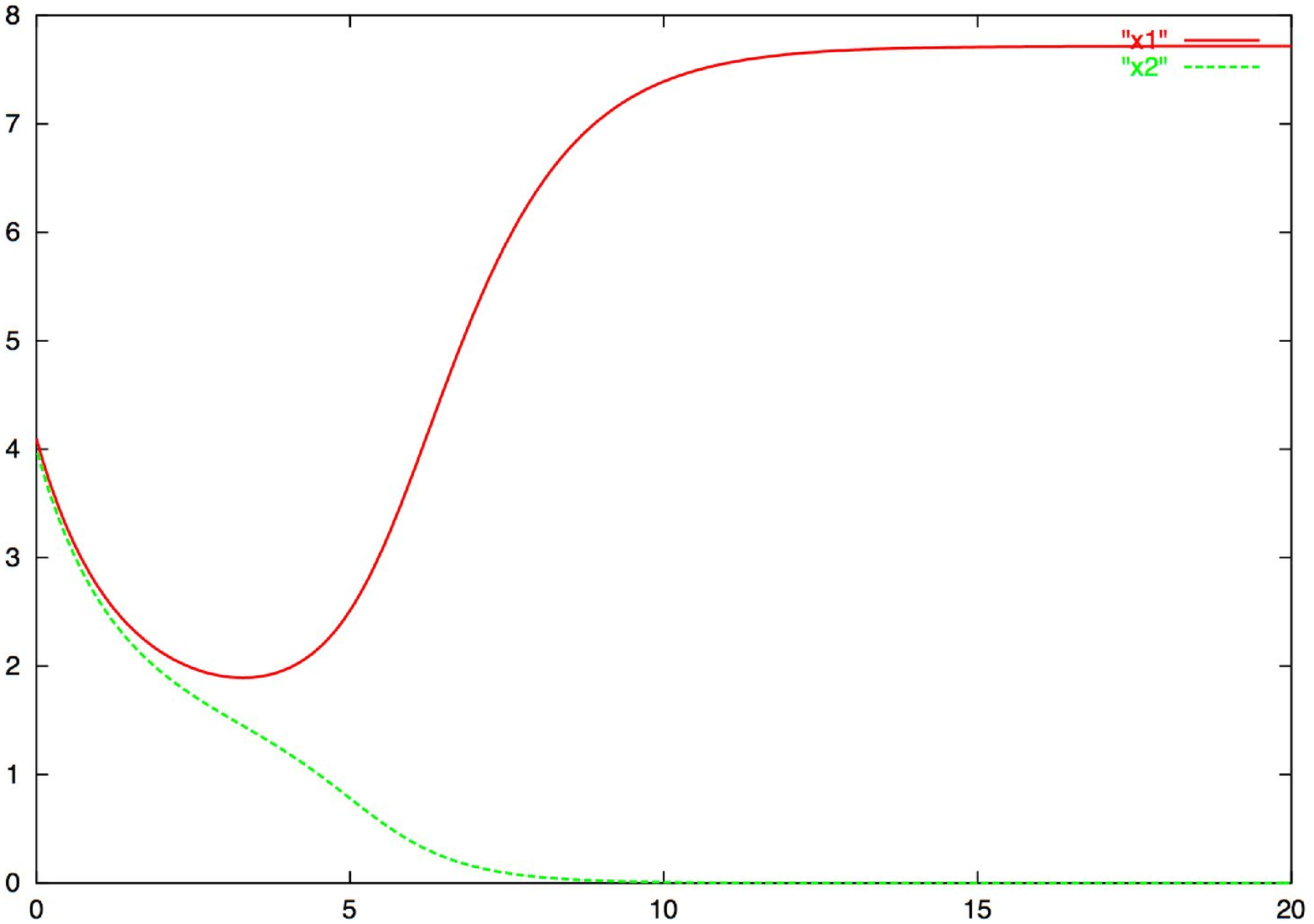} \\
  \mathbf{b}
\end{array}$
\caption{Same as Figure \ref{ic_resting}, but with initial concentrations at roughly half the equilibrium value when one element is on and all others off. Initial concentrations differ by 0.5 ($\mathbf{a}$), or 0.1 ($\mathbf{b}$).}
\label{ic_half_resting}
\end{center}
\end{figure}

To show the effect in more detail, individual trajectories were plotted for a 2-dimensional switch (Figures \ref{ic_resting} and \ref{ic_half_resting}). For cell fusion and reprogramming experiments, the effect on the concentration of switch elements depends on the dynamics of nuclear import and export. Two types of initial conditions were used: two switch elements were given the concentration that one would have at rest, if it was "on" (Figure \ref{ic_resting}), or two switch elements were given half that concentration (as cytoplasmic concentrations of proteins expressed exclusively in 1 of 2 equally-sized cells should be cut by half upon fusion; Figure \ref{ic_half_resting}). In both cases, the concentrations of the two switch elements, even for that which will eventually prevail, initially go down. The effect is more pronounced for higher values of the initial concentrations, and for close initial values of the two concentrations. This could explain the transient extinction of expression of differentiated markers upon cell fusion (see Discussion): if there is sufficient cooperativity downstream of the switch element, its dip could be sufficient to provoke a temporary extinction of expression of the proteins specific to the differentiated state.

\subsubsection{Extinction domain}

For $\alpha>\frac{\sigma^{2}}{4\left(n\sigma +1 \right)}$, there is an extinction domain around the diagonal $x_1=..=x_n$. Simulations show that the domain is very restricted until $\alpha$ becomes very close to its upper threshold value, at which non-0 equilibria cease to exist (see Figures \ref{alpha15} and \ref{alpha24}).

\subsubsection{Summary of $\alpha$ threshold values}
\label{alpha_summary}
For the system to be able to sustain $k$ switch elements "on" at the same time, the condition $\alpha<1/k^2$ must be met (for $\sigma \gg 1$, this condition is also sufficient). Thus, for $\alpha>1/4$, only 1 switch element can be on at a time. The corresponding equilibrium value is a decreasing function of $\alpha$. For $\alpha>\frac{\sigma^{2}}{4\left(n\sigma +1 \right)}$, there is an "extinction domain" around the diagonal $x_1=..=x_n$: matching sufficiently closely the concentrations of the switch elements, at whatever value, makes the system switch off all switch elements. For large $\sigma$, the extent of this domain is small, except in a very narrow range of $\alpha$ values. Finally, for $\alpha>\frac{\sigma^2}{4\left( \sigma +1 \right)}$, a condition which becomes $\alpha > \sigma/4$ for large $\sigma$, there are no non-0 steady states.

\section{Discussion}

\subsection{Co-expression properties}
Of the models proposed here, if the cooperativity of activation is considered to be constant, only the model with bHLH dimerisation is capable both of exclusive expression of an arbitrary number of switch elements, and coexpression at similar levels of all elements. This behaviour is finely tunable with the key competition parameter deriving from the availability of the bHLH hetero-dimerisation partner, with lower availability being unfavourable to co-expression of the antagonistic factors (see below for a further discussion).

The model with mutual inhibition, autocatalysis, and leak can express no more than one switch element at a level higher than the other ones, and is thus less amenable to progressive elimination of unwanted factors in the course of differentiation. In order for coexpression to occur at a significantly-higher level than the leak, the cooperativity of the system must be close to 1. If differentiation was controlled by a network of this kind, initial coexpression could be maintained either by a low transcriptional strength in the system (which is consistent with antagonistic factors being expressed at a lower level in the un-differentiated state), or, as has been suggested, by regulated degradation of mRNAs.

Interestingly, the multistability behaviours of a switch based on bHLH-like dimerisation and that of a switch based on direct cross-repression are qualitatively different: the former can maintain many variables on at an equilibrium only if those variables are sufficiently high (compared to the transcription strength), while the reverse is true of the latter.

We previously studied networks of cross-repressing factors, in which the factors do not enhance their own expression \citep{CinquinO:02}. We did not include this kind of model in the present study, because for one factor to be able to dominate all the others, it had to be assumed that the cooperativity of the network was very high, an assumption which is possibly not realistic.

\subsection{Peaks of differentiation inhibitors}

According to the paradigm of inhibition of differentiation by sequestration of class A bHLH proteins, the quantity of binding partner should be low prior to differentiation, and the competition parameter $\alpha$ introduced earlier should thus be high. Relieving inhibition of differentiation, by increasing the quantity of binding partner, and thus decreasing competition, cannot move the bHLH dimerisation network from a state where it supports coexpression of many switch elements, to a state where only one is expressed. Also, increasing the transcription strength of the network $\sigma$ does not destabilise equilibria.

It is thus impossible to account for the selection of one differentiation outcome by increasing the availability of class A proteins (for example by downregulation of Id proteins). However, it is still possible that the competition strength, even in the undifferentiated state, is sufficiently low for many switch elements to be co-expressed. A potential mechanism to select 1 element and extinguish all others is then to transiently increase the competition strength, for example by transient up-regulation of Id proteins, just prior to differentiation commitment (a corresponding simulation is shown in Figure \ref{bell_alpha}). This is in fact what has been experimentally observed on independent occasions, on a short time scale, during in vitro differentiation of osteoblasts \citep{OgataT:93}, neurons \citep{NagataY:94}, myeloid cells \citep{IshiguroA:96}, astrocytes \citep{Andres-BarquinPJ:97}, Schwann cells \citep{StewartHJ:97}, keratinocytes \citep{LanglandsK:00}, germ cells \citep{HouldsworthJ:01}, and fibroblasts \citep{ChambersRC:03}. No rationale for these transient effects had been proposed so far.

When Id proteins are not up-regulated, other proteins could play the same role of increasing competition in the bHLH network. For example, Hes-1, which also has class A sequestering activity \citep{SasaiY:92}, is transiently upregulated upon differentiation of an immortalised hair cell line \citep{RivoltaMN:02}, an immortalised neural cell line \citep{OhtsukaT:98}, and neuroblastoma \citep{GrynfeldA:00} (although its role in the latter case is complicated by the fact that it also binds Id proteins and interferes with Id2/E2-2 complex formation, \citealp{JogiA:02}); the transient Hes-1 expression is concomitant with downregulation of the bHLH protein HASH-1. Hes genes are dominant repressors with many targets \citep{BaroloS:97}, and could also directly repress many elements of the network, which can be modeled by a decrease in the transcription strength $\sigma$, and has the same effect of destabilising equilibria where many elements are coexpressed.

Finally, erythroid-specific genes have been observed to be transiently downregulated upon induced, in-vitro differentiation \citep{ListerJ:95}, which could also be explained by transiently-increased competition in a bHLH dimerisation network.

\subsection{Dynamical properties}
\label{dynamics}
Analysis of the proposed dynamical systems shows that the time to convergence can widely depend on the initial condition. Convergence can be relatively very slow when initial conditions are near a threshold around which the system converges to two or more different outcomes. This is for example the case when 2 or more "switch elements" are at roughly equal concentrations, higher than that of others (the more elements in competition, the slower the competition becomes). It is interesting to note that slow effects are observed in induced-transdifferentiation experiments, and in cell fusion experiments. 

\begin{itemize}

\item{Fibroblasts reprogrammed to T-cell-like cells need to be incubated for many days before they acquire detectable T-cell characteristics \citep{HakelienAM:02}. This may be due to the fact that fibroblast master genes are expressed at a high level, and the counter-acting T-cell master genes, introduced by permeabilisation of the membranes, are also present at a high concentration. An effect of the relative levels of cytoplasmic factor concentrations could be tested by incubation in T-cell and fibroblast cytoplasmic extracts, mixed at different ratios. Further investigation of master networks could involve incubation of cells in cytoplasmic extracts of 3 or more cells-types (or transient misexpression, at controlled levels, of antagonistic master genes).}
\item{In hepatoma-fibroblasts hybrids, extinction of albumin production can take days \citep{Mevel-NinioM:81}. Most interestingly, some hybrids show reexpression of albumin after extinction. These two outcomes can be accounted for by the models proposed above: when two antagonistic "switch elements" are coexpressed at a high level (which probably corresponds to the fusion experiments, as upon fusion the protein contents of the cells, which are of different phenotypes, are mixed), it is possible for the system to revert to a state where all switch elements are turned off (total extinction), or for the two switch elements to decrease to a low level, before the trajectory of one of them picks up and goes back to a high state (extinction followed by reexpression). }
\item{Activation of the myogenic phenotype also takes place on the scale of days, when muscle cells are fused to various other cell types, a delay which was suggested not to be linked to DNA duplication (\citealp{BlauHM:85}; see \citealp{BlauHM:99}, for an extensive review).}
\end{itemize}

Also, it could be that the progressive upregulation of differentiation-related genes observed during hematopoietic development is a cell-autonomous consequence of the slow dynamics of a switch network.

\subsection{Stochastic outcomes}

It has been observed in the studies cited above that heterokaryons with the same number of nuclei coming from each donor can have different differentiation responses. \citet{BlauHM:99} suggested that the differentiation outcomes are regulated by a network of opposing factors. Within this framework, stochastic responses to cell fusion can readily be explained by slight differences in the quantities of differentiation factors contributed by each cell type, which tip the balance one way or the other. The network determining cell fate would be most sensitive to random noise when the factors are in roughly equal concentrations. The sensitivity to molecular noise of the networks proposed here would be interesting to study, as it has been proposed that some types of differentiation during development could have a stochastic aspect (for example in adipogenesis, \citealp{TchkoniaT:02}, or hematopoiesis, \citealp{EnverT:98}). 

\subsection{Evolvability of switch networks}

In addition to having a coexpression behaviour which is easily tunable by one parameter, the generic bHLH network studied here has the advantage of being perhaps more easily evolvable than a network in which every element actively represses all others: the interaction needs only take place between every element and a common activator (which requires $n$ interactions, instead of $n \left(n-1\right)/2$). bHLH networks have been suggested to evolve mainly by single-gene duplication events \citep{AmoutziasGD:04}, maintaining a topology in which every element of the network interacts with a restricted number of "hubs".

\section{Conclusion}

The models presented here could be useful in understanding development, as well as cell-fate reprogramming (which can be induced artificially, but has also been shown to happen naturally, \citealp{WeimannJM:03}). We have derived general results about the dynamics and co-expression properties of switch networks, and shown the flexibility of bHLH dimerisation networks. Of the networks studied here, which were generically formulated with usual kinetic equations, only a subset can co-express antagonistic elements at a similar level, higher than the basal level: those with mutual inhibition, autocatalysis, and leak (but only when the cooperativity is very close to 1, and the transcription strength sufficiently low), and bHLH dimerisation networks (when the competition is sufficiently weak). This restricts the classes of models which can reproduce experimentally-observed co-expression of antagonistic factors, as well as showing how it can occur.

Strikingly, even though bHLH networks are the most apt to coexpression of antagonistic elements, the selection of one element requires a transient increase in competition, which is not what is thought to happen over a long time scale in the course of differentiation. Transient, hitherto-unexplained increases in competition have however been observed in a few cell lines upon differentiation, and could be a general phenomenon.

In order to model specific differentiation events, these networks would probably need to be extended to take into account combinatorial interactions, which could complicate their behaviour.  The models would also gain from being extended to take into account non-symmetrical networks, in which some switch elements are stronger than others, and stochastic kinetics.

\section*{Acknowledgements}

This work was funded by an AstraZeneca studentship awarded to OC by CoMPLEX.

\bibliography{references,sup_references}

\appendix

\section{Analysis of mutual inhibition with autocatalysis}
\label{mia}

\subsection{Special case: no cooperativity ($c=1$)}

We assume that $\sigma>1$. The set of steady states for the system defined by equations \ref{equations_autocat} is 0 and the attracting hyperplane $\{x\ |\ 1 + \Sigma_{i=1}^{n} x_i=\sigma\}$. Let $s=\Sigma_{i=1}^{n}x_i$. Then $s$ never crosses the value $\sigma-1$, and since $\dot{x_i}=x_i \left( \frac{\left(\sigma -1\right)-s}{1+s} \right)$, $\dot{x_i}$ is of constant sign, and each $x_i$ convergent.

 Simulations show that there is a great number of stable steady states.

For $c>1$, the convergence of the dynamical system (defined by equations \ref{equations_autocat}) to an equilibrium, from any initial condition, will be derived in a more general context, in section \ref{convergence_autocat_leak}. In the rest of the appendix we assume $c>1$.

\subsection{One on, all others off}

\subsubsection{Equilibrium existence}

The steady-state equations are

$$\forall j, \bar{x}_{j} \left( 1+\Sigma_{i=1}^{n}\bar{x}_{i}^\mathrm{c} \right)=\sigma \bar{x}_{j}^{c},
$$
ie
$$
\bar{x}_{j}^{c-1}=\frac{1}{\sigma}\left( 1+\Sigma_{i=1}^{n}\bar{x}_{i}^\mathrm{c} \right) \ \mathrm{or}\ \ \bar{x}_j=0
$$

Re-arranging the first equation,

$$
\frac{1}{\sigma}\bar{x}_j^c -\bar{x}_{j}^{c-1}=- \frac{1}{\sigma}\left( 1+\Sigma_{i\ne j}\bar{x}_{i}^\mathrm{c} \right)
$$

Let $f(x)=\frac{1}{\sigma}x^c -x^{c-1}$. Then $f'(x)=\frac{c}{\sigma}x^{c-1}-(c-1)x^{c-2}$. $f'(x)<0$ iff $\frac{c}{\sigma}x<c-1$. The minimal value of $f$ over the positive real set is $f(\frac{c-1}{c}\sigma)=\frac{1}{\sigma}\left( \frac{c-1}{c}\sigma \right) ^{c} - \left( \frac{c-1}{c}\sigma \right) ^{c-1}=\sigma^{c-1} \left(  \frac{c-1}{c}  \right) ^{c-1} \left( \frac{c-1}{c} - 1 \right)=\sigma^{c-1} \left(  \frac{c-1}{c}  \right) ^{c-1} \frac{-1}{c}$.

The equilibria studied here are such that only $\bar{x}_j$ is non-0, for some $j$. There are either 0 or 2 solutions, 2 iff

$$
\sigma^{c-1} \left(  \frac{c-1}{c}  \right) ^{c-1} \frac{1}{c} > \frac{1}{\sigma}
$$
$$
\sigma^{c} > c \left( \frac{c}{c-1} \right) ^{c-1}
$$

\begin{equation}
\label{iff_existence_one_on}
\ln{\sigma} > \frac{\ln{c}+\ln{\left( \frac{c}{c-1} \right) ^{c-1}}}{c}
\end{equation}

$\ln{\left( \frac{c}{c-1} \right) ^{c-1}}$ is an increasing function of $c$, and $\lim_{c \to \infty} \ln{\left( \frac{c}{c-1} \right) ^{c-1}} =1$. $\frac{ln{c}}{c}$ is decreasing for $c>e \simeq 2.7$. The right-hand side of equation \ref{iff_existence_one_on} has a maximum for $c=2$, of about $0.7$, matched by $\sigma=2$. Thus, for  $\sigma \ge 2,$ there are two equilibria. Both large $c$ and large $\sigma$ are favourable to the existence of an equilibrium with one variable dominating all others.

\subsubsection{Local stability analysis}

It is useful, for the Jacobian term computations to follow in the rest of the appendix, to note that if $g(x)=\frac{x^m}{\alpha+x^m}$, $g'(x)=\frac{\alpha m x^{m-1}}{\left( \alpha + x^m \right) ^2}$.

If $x_j$ is at a non-zero steady-state and $\forall i\ne j, x_i=0$, and if $c>1$, the stability at that steady state depends only on the sign of the $(j,j)$ coefficient of the Jacobian matrix (this coefficient will be called $J_{j,j}$ in the remainder of the appendix).

\begin{equation}
\label{jac_auto}
J_{j,j}= -1 + \sigma c \left( 1+ \Sigma_{i \ne j} x_{i}^{c} \right) \frac{x_{j}^{c-1}}{\left( 1+\Sigma_{i=1}^{n}x_{i}^\mathrm{c} \right)^2}
\end{equation}

$$
J_{j,j}= -1 + \sigma c \frac{x_{j}^{c-1}}{\left( 1+x_{j}^\mathrm{c} \right)^2},
$$

with

\begin{equation}
\label{m}
\sigma \bar{x}_{j}^{c-1}=1+\bar{x}_j^{c},
\end{equation}
at equilibrium
$$
J_{j,j}= -1 +  c \frac{1}{\sigma \bar{x}_{j}^{c-1}},
$$

the equilibrium is stable iff

$$
\bar{x}_{j}^{c-1}>\frac{c}{\sigma},\ \mathrm{ie}\ 1+\bar{x}_{j}^{c}>c
$$

It is possible to give a sufficient condition for the equilibrium with the greatest solution to equation \ref{m} to be stable. Let $f(x)=x^{c} - \sigma x^{c-1}$. If $f\left( \left( \frac{c}{\sigma} \right) ^ {\frac{1}{c-1}} \right) < -1$, then the greatest root of equation \ref{m} will be greater than $\left( \frac{c}{\sigma} \right) ^ {\frac{1}{c-1}}$, and the corresponding equilibrium will be stable. A sufficient stability condition is thus

$$
\left( \frac{c}{\sigma} \right) ^ {\frac{c}{c-1}} <c-1
$$

Numerical investigation shows that this condition is met for $\sigma \ge 2$.

\subsection{$k$ variables on, others off}

With identical parameters, there can be no equilibrium with 2 variables having different, non-zero values.

At any equilibrium, variables can be renumbered so that, in the Jacobian matrix, variables at 0 form an independent block. This block is stable, and the stability of the whole system depends only on the block formed by non-0 variables. Thus, in the following we suppose that no steady-state variable has 0 for a value.

For $i \ne j$, 
$$J_{i,j}(\bar{x})=-\sigma c \frac{\bar{x}^{2 c -1}}{\left( 1 + k \bar{x} ^{c} \right)^2}$$

With the same kind of analysis as in \citet{CinquinO:02}, the equilibrium is stable only if 

{\large
\label{f1}
\begin{align}
\nonumber  \sigma c \frac{\bar{x}^{2 c -1}}{\left( 1 + k \bar{x} ^{c} \right)^2}  <  1 - \\
 \sigma c \left( 1 + \left( k - 1 \right) \bar{x}^{c} \right) \frac{\bar{x}^{c-1}}{\left( 1 + k \bar{x}^{c} \right) ^2}    
\end{align}
}

With the definition of the equilibrium,

$$
\sigma c \bar{x}^{2c-1} < \sigma ^2 \bar{x}^{2c-2} -\sigma c \left( 1 + \left(k-1 \right)\bar{x}^c\right)\bar{x}^{c-1}
$$

$$
x^c  < \frac{\sigma}{c}x^{c-1} - \left( 1 + \left( k -1 \right) x^c \right)
$$

$$
\frac{\sigma}{c} x^{c-1} > k x^c +1
$$

Again with the definition of the equilibrium,

$$
\frac{1}{c}x^{c-1}>x^{c-1},
$$

ie $c<1$, in which case no interesting equilibria exist.

\section{Analysis of mutual inhibition with autocatalysis, and leak}
\label{mial}

If $\alpha\ge 0$, $c\ge 1$, and one of these inequalities is strict, the function $f(x)=x^{1-c}-\alpha x^{-c}$ can take the same value for at most 2 positive values of $x$. Thus, there are only two values a variable can take at a given steady state (0 cannot be a steady state value). If two different equilibrium values are taken by some variables, one of these values is higher than $\alpha \frac{c} {c-1}$, and the other lower.

If $\alpha>0$ and $c=1$, the system only has one equilibrium, with all variables equal.

\subsection{Convergence}
\label{convergence_autocat_leak}
Let $y_{i}= \sqrt{x_i}$, and

\begin{align}
\nonumber P=\frac{1}{4}\Sigma_{i=1}^{n} y_{i}^2 - \frac{\sigma}{4c}\log{\left( 1+\Sigma_{i=1}^{n}y_{i}^{2c} \right)} -\\
\nonumber  \frac{1}{2}\log{\Pi_{i=1}^{n}y_{i}^{ \alpha}} 
\end{align}

$$
\dot{y_i}=\frac{\dot{x_i}}{2\sqrt{x_i}}
$$

$$
2\dot{y_i}=- y_i + \sigma  \frac{y_{i}^{2c-1}}  {1+\Sigma_{i=1}^{n} y_{i}^{2c}} + \frac{ \alpha}{y_i}=2\frac{\partial P}{\partial y_i}
$$

Thus, $P$ is a potential for the system.

If its equilibria are isolated, a gradient system converges to a steady-state regardless of the initial conditions. It is shown below that the number of solutions of the system is finite when the cooperativity $c$ is an integer, and the system thus always converges to a steady state (we expect this result to also hold for non-integer values of $c$). The model without leak corresponds to $\alpha=0$, and this convergence result thus also applies to it, for $c>1$.

\subsection{Steady-state analysis: all at the same value}

\subsubsection{Equilibrium existence}

$$\forall\ j,\  \left(x_{j} - \alpha \right) \left( 1+\Sigma_{i=1}^{n}x_{i}^\mathrm{c} \right)=\sigma x_{j}^{c}
$$

If $\forall\ j,\ x_{j} = \bar{x}$, 

\begin{equation}
\label{ss_all_eq_leak}
n \bar{x}^{c+1} - \left( \sigma + n \alpha \right) \bar{x}^c +\bar{x}-\alpha=0
\end{equation}

There is at least one solution, maybe 3 (or 2 in degenerate cases) depending on the parameters. The solutions are noted $\bar{x}_l$, $\bar{x}_u$, and $\bar{x}_h$, with $\bar{x}_l<\bar{x}_i<\bar{x}_h$. 

If $f(x)=n x^{c+1} - \left( \sigma + n \alpha \right) x^c +x$, $f'(x)=(c+1) n x^c - c (\sigma + n \alpha)x^{c-1} +1$, $f''(x)=c(c+1)nx^{c-1} - c(c-1)(\sigma+n\alpha)x^{c-2}$. $f''\left(\frac{c-1}  {n(c+1)} \left(\sigma + n\alpha\right)\right)=0$.
$f'$ takes negative values iff $f'\left(\frac{c-1}  {n(c+1)} \left(\sigma + n\alpha\right)\right)<0$, which is a necessary condition for the existence of 3 equilibria with all variables on.

The dynamics of the system constrained to $\forall\ i,\ x_i=x$ are defined by

$$
\dot{x}=-x + \frac{\sigma x^c}{1+nx^c} + \alpha
$$

The sign of $\dot{x}(t)$ is the opposite of that of $f(x(t))$. Because $\bar{x}_u$ is such that $f'(\bar{x}_i)<0$, it is easy to see that the steady state $\bar{x}_u$ is unstable for the constrained system, and thus for the full system.

\subsubsection{Local stability analysis}

With a leak $\alpha$, equation \ref{f1} becomes

$$
\bar{x}^c < \frac{\sigma}{c} \frac{\bar{x}^{c+1}}{\left( \bar{x}-\alpha\right)^2} -1 -\left(k -1 \right) \bar{x}^c
$$

$$
1+k \bar{x}^c<\frac{\sigma}{c} \frac{\bar{x}^{c+1}}{\left(\bar{x}-\alpha \right)^2}
$$

$$
\frac{\sigma \bar{x}^c}{\bar{x}-\alpha} < \frac{\sigma}{c} \frac{\bar{x}^{c+1}}{\left(\bar{x}-\alpha \right)^2}
$$

$$
\bar{x}-\alpha < \frac{1}{c}\bar{x}
$$

Thus the stability condition \ref{f1} is met iff $\bar{x}<\alpha \frac{c}{c-1}$ (in that case, since non-diagonal terms of the Jacobian are obviously negative, diagonal terms are also negative, and the equilibrium is stable). Since solutions to equation \ref{ss_all_eq_leak} can be made arbitrarily high by increasing $\sigma$, increasing $\sigma$ past a threshold value (other parameters being equal) will prevent the existence of a stable equilibrium with all variables equal.

\subsection{$k$ on, $k<n$}
\label{kon_kppqn}
Let $p=n-k$.

$$
\left(\bar{x}_l-\alpha\right)\left( 1+p\bar{x}_{l}^c + k\bar{x}_{h}^c \right) = \sigma \bar{x}_{l}^c
$$

\begin{eqnarray}
\nonumber
 p \bar{x}_{l}^{c+1} - \left( p\alpha +\sigma \right) \bar{x}_{l}^c & + &  \left( 1+k \bar{x}_{h}^c \right)\bar{x}_{l} -\\
 \nonumber \alpha \left( 1+k \bar{x}_{h}^c \right) & = & 0 \\
 \nonumber
k \bar{x}_{h}^{c+1} - \left( k\alpha +\sigma \right) \bar{x}_{h}^c & + & \left( 1+p \bar{x}_{l}^c \right)\bar{x}_{h} -\\
\nonumber \alpha \left( 1+p \bar{x}_{l}^c \right) & = & 0
\end{eqnarray}

Choosing for example the graded lexicographic order over $\mathbb{C}[{x}_{l},{x}_{h}]$, theorem 5.3.6 from \citet{ideal_var_alg} shows that the system has a finite number of solutions, when $c$ is an integer.

We have

$$
J_{i,i}=-1 + c\sigma x_{i}^{c-1}\frac{D-x_i^c} {D^2}
$$

$$
J_{i,j}=-c\sigma x_j^{c-1}\frac{x_i^c}{D^2}
$$

If $x_i=x_j$,

$$
J_{i,i}-J_{i,j}=-1 + c x_i^{c-1} \frac{\sigma}{D}=-1 + c\frac{x_i - \alpha}{x_i}
$$

Consider the reordered Jacobian matrix, with $k$ variables "on" with a value $\bar{x_h}$, and $p$ "off" with a value $\bar{x_l}$ ($k+p=n$).

It follows from the analysis in section \ref{on_diff_val} that the equilibrium can be stable only if $J_{i,i}-J_{i,j}<0$ (ie $x_i<\alpha \frac{c}{c-1}$), if the number of variables having value $x_i$ is strictly greater than 1.

Thus there are only two possible kinds of stable equilibria: all variables equal, in which case the equilibrium value is lower than $\alpha \frac{c}{c-1}$, or one higher than all the other ones (in which case the lower ones are lower than, and the higher one greather than $\alpha \frac{c}{c-1}$).

\section{Analysis of the bHLH model}
\label{bHLH_analysis_details}

Without cooperativity in transcriptional activation by the bHLH dimer, there is only one stable steady-state:

$$
\dot{x_i}=x_i \left( -1 +\frac{\sigma}{\alpha \left( 1+\Sigma_{j=1}^n x_j \right) +x_i} \right)
$$

If at some steady state $k$ variables are on and share a common value $\bar{x}$ (variables at a steady state, if not 0, share a common value),

$$
1=\frac{\sigma}{k\alpha \bar{x}+\bar{x} +\alpha}
$$

$$
\bar{x}=\frac{\sigma-\alpha}{k\alpha+1},
$$

and if $x_p(t_0)=0$,

$$
J_{p,p}=\left( -1 + \frac{\sigma}{k\alpha \bar{x} + \alpha} \right)
$$

$$
J_{p,p}=\frac{\sigma-\alpha}{\alpha \left( k\sigma +1 \right)}>0,
$$

and $J_{p,l}=0$ for $p \ne l$, proving the unstability of the steady state.

In the following, it is assumed that transcriptional activation occurs with cooperativity 2, and the steady-state equations become

$$
\bar{x}_i=\sigma  \frac{\bar{x}_i^2} {\left(\frac{D}{a_t}\right)^2 K_2 + \bar{x}_i^2}
$$

\begin{equation}
\label{bhlh_ss}
\forall\ i,\ \alpha D^2 + \bar{x_i}^2=\sigma \bar{x_i}
\end{equation}

\subsection{Dynamical analysis}

0 is a stable steady state. If $x_i(0)=0$, then $\forall\ t>0,\ x_i(t)=0$. If $x_i(0)>0$, then $\forall\ t>0,\ x_i(t)>0$. One can thus suppose that $\forall\ i,\forall\ t\ge 0,\ x_i(t)>0$.
Consider a state in which there is one variable strictly superior to all others (ie, a state not belonging to the line $x_1=x_2=..=x_n$). Suppose without loss of generality that the variable in question is $x_1$. Consider the function

$$
f_{1}(x)=\frac{x_1^2}{\alpha D^2 + x_1^2}
$$

$$
\dot{f_{1}(x)}=2\alpha D x_1\frac{\dot{x_1}\left( D-x_1 \right) - x_1 \Sigma_{i=2}^n \dot{x_i}}{\left( \alpha D^2 + x_1^2 \right)^2}
$$

\begin{align}
\nonumber \frac{\left( \alpha D^2 + x_1^2 \right)^2}{2\alpha D x_1}\dot{f_{1}(x)}=\dot{x_1} +\\
\nonumber  \sigma \Sigma_{i=2}^n \frac{x_1 x_i \left( x_1 - x_i \right)\left( \alpha D^2 - x_1 x_i \right)}  {\left(\alpha D^2 + x_1^2 \right)  \left( \alpha D^2 + x_i^2 \right)}
\end{align}

For $\alpha \ge 1/2$, the second term is positive.

We have

$$
\frac{\mathrm{d}x_1(t)}{\mathrm{d}t}= \sigma f_{1}(x) - x_1
$$

We first consider the case in which $\forall\ t\ge 0,\forall\ n\ge j>1,\ x_1>x_j$.

Suppose that $\sigma f_{1}(0) \ge x(0)$. In this case, $\dot{f_1}\left( 0 \right) > 0$, and $x_1$ and $f$ are strictly increasing functions of time. If $\sigma f_{1}(0)<x_1(0)$, then $\dot{f_1}\left( 0 \right)$ can be negative or positive. In the first case, $x_1$ is decreasing as long as $f_1$ is.
If at some time $t_0$ $\sigma f_{1}(t_0) \ge x(t_0)$, then for $t>t_0$, $x_1$ and $f_1$ are increasing functions of time. Thus there can be at most one change in the monotony of $x_1$.  Thus $\lim_{t\rightarrow \infty} x_1(t)$ exists. Since $\ddot{x_1}$ exists and is bounded on any trajectory, $\lim_{t\rightarrow \infty} \dot{x_1}(t)=0$. All trajectories thus converge to a steady state where $\forall j>1,\ x_j=x_1\ \mathrm{or}\ x_j=0$.

If $\exists\ t\ \mathrm{st}\ \forall \ n>j>1,\ x_1(t)=x_j(t)$, the system is brought back to one dimension. Note that it is impossible for any variable to outgrow $x_1$.

\subsection{Steady-state analysis: variables on at the same value}
\subsubsection{Equilibrium existence}

Variables zero at the steady state can be discarded from the analysis. If $k$ variables are non-0, and  are all equal, to $\bar{x} \ne 0$,

\begin{equation}
\label{bhlh_1}
\bar{x}^{2}\left(1 + k^2\alpha \right) + \bar{x}\left( 2k\alpha - \sigma \right) + \alpha = 0
\end{equation}

Solutions are

$$
\frac{\sigma -2k\alpha \pm \sqrt{\sigma^2-4\alpha\left(1+k\sigma\right)}}{2\left(1+k^{2} \alpha\right)}
$$

A sufficient and necessary condition for the existence is

$$
4 \alpha \frac{k\sigma + 1}{\sigma^2} < 1
$$

It will be shown below that, at a stable steady-state, there is at most 1 non-0 variable which can be different from other non-0 variables. If there is such a variable, equal to $y$, the equation for the value of other variables becomes

\begin{align}
\label{bhlh_2}
\nonumber \bar{x}^{2}\left(1 + k^2\alpha \right) + \bar{x}\left( 2k\alpha \left( 1+y \right) - \sigma \right) +\\
 \alpha \left(1 + y \right)^2= 0
\end{align}

Solutions are

$$
\frac{\sigma -2k\alpha\left(1+y\right) \pm \sqrt{\sigma^2-4\alpha\left(1+k\sigma+y\right)\left(1+y\right)}}   {2\left(1+k^{2} \alpha\right)}
$$

and the condition for a solution to exist 

$$
4 \alpha \left(1+y \right) \frac{k\sigma + 1+y}{\sigma^2} < 1
$$

The solutions for $y$ are

$$
\frac{\sigma -2\alpha\left(1+k\bar{x}\right) \pm \sqrt{\sigma^2-4\alpha\left(1+\sigma+k\bar{x}\right)\left(1+k\bar{x}\right)}} {2\left(1+\alpha\right)}
$$

\subsubsection{Local stability analysis}

Variables zero at the steady state can be discarded from the analysis.

Using 

$$
\dot{\frac{x^2}{ax^2+bx+c}}=\frac{bx^2 +2cx}{\left( {ax^2+bx+c} \right)^2},
$$

one derives the diagonal term of the Jacobian (with $b=2\alpha\left(D-x_i\right)$ and $c=\alpha \left(D-x_i\right)^2$):

$$
J_{i,i}=-1 + 2 \sigma \alpha x_i \frac{D \left(D-x_i \right)}{\left( \alpha D^2 + x_i^2 \right)^2}
$$

Using the steady state equation \ref{bhlh_ss},

\begin{align}
\nonumber J_{i,i}=-1 + \frac{2 \alpha}{\sigma x_i}  \left( D \left(D-x_i \right)\right)=-1 +\\
\nonumber \frac{2}{\sigma}\left( \sigma -x_i - \alpha D \right)
\end{align}

$$
J_{i,i}=1-\frac{2 }{\sigma} \left( x_i+\alpha D  \right)=1-\frac{2}{\sigma}\left( \alpha + x_i\left(1+k\alpha \right) \right)
$$

The diagonal terms are negative for

$$
x_i>\frac{\sigma / 2 -\alpha}{1+k\alpha}
$$

The off-diagonal terms are given by

$$
J_{i,j}=- \sigma  x_{i}^2 \frac{ 2\alpha x_j + 2 \alpha \left( D - x_j \right)}{\left( \alpha D^2 + x_i^2 \right)^2}
$$

$$
J_{i,j}=- 2 \sigma \alpha x_{i}^2 \frac{ D}{\left( \alpha D^2 + x_i^2 \right)^2}
$$

$$
J_{i,j}= -\frac{2\alpha}{\sigma}D
$$

$$
J_{i,j}-J_{i,i}=-1+2\frac{x_i}{\sigma}
$$

Thus, a necessary condition for the equilibrium to be stable is 

\begin{equation}
\label{bhlh_stab_cond}
\forall\ \bar{x}_i\ \mathrm{st}\ \bar{x}_i\ne 0,\ \bar{x}_i>\sigma/2
\end{equation}

This is possible if and only if {$\alpha<1/k^2$} and $\sigma>2\frac{k\alpha+\sqrt{\alpha}}{1-k^{2} \alpha}$.

Condition \ref{bhlh_stab_cond} is stronger than the requirement for the diagonal element to be negative (and is thus also a sufficient condition), and can never be met by variables equal to the lower solution of equations \ref{bhlh_1} or \ref{bhlh_2} .

Thus, for any value of the transcription strength $\sigma$ and for any number of coexistant variables $k$, sufficiently low values of $\alpha$ make the equilibrium stable. If there is a stable equilibrium with $k$ variables on, there is also a stable equilibrium with $p$ variables on, for $1<p<k$. For sufficiently large $\sigma$, the necessary condition $\alpha<1/k^2$ becomes sufficient for stability (see Figure \ref{increasing_alpha} for an illustration of the validity of this condition).

\subsection{On at different values}

\label{on_diff_val}
If at steady state, $x_i \ne x_j$ and both are non-0, then

$$
x_i^2-\sigma x_i = x_j^2 -\sigma x_j \left( =-\alpha D^2 \right)
$$ 

There are thus only two possible non-0 steady-state values, noted $\bar{x}_a$ and $\bar{x}_b$, with $\bar{x}_a<\bar{x}_b$. Noting $P(x)=x^2-\sigma x$, and supposing that $\bar{x}_a$ and $\bar{x}_b$ exist, $P'(\bar{x}_a)<0$, \emph{ie} $\frac{2 \bar{x}_a}{\sigma}<1$.

Consider the Jacobian matrix of the system, reordered so that variables having $\bar{x}_a$ as a value come before those having $\bar{x}_b$ as a value:

$$
\label{ss_2matrix}
\begin{pmatrix}
  \overbrace{\begin{matrix}  a_{\phantom{2}}    &  c & \cdots &  c \\ c_{\phantom{2}}&  \ddots & \cdots & \vdots \\ \vdots & \vdots& \ddots & \vdots \\ c_{\phantom{2}} & \cdots & c & a \end{matrix}}^{k} & \overbrace{\begin{matrix} f_1 & \cdots & \cdots & f_1 \\  \vdots & \vdots &  \vdots & \vdots  \\ \vdots & \vdots &  \vdots & \vdots \\  f_1 & \cdots & \cdots & f_1 \end{matrix}}^{p} \\
  
  \begin{matrix} f_2 & \cdots & \cdots & f_2 \\  \vdots & \vdots &  \vdots & \vdots  \\ \vdots & \vdots &  \vdots & \vdots \\  f_2 & \cdots & \cdots & f_2 \end{matrix} & \begin{matrix} b    &  e & \cdots &  _{\phantom{1}} e \\ e  &  \ddots & \cdots & \vdots \\ \vdots & \vdots& \ddots & \vdots \\ e & \cdots & e & _{\phantom{1}} b \end{matrix}
\end{pmatrix}
$$

With the appropriate eigenvectors, it is easy to show that $b-e$ and $a-c$ are eigenvalues for this matrix, of order $k-1$ and $p-1$. Thus, if $k>1$ and $p>1$, a necessary condition for stability of an equilibrium is $e>b$ and $c>a$. In particular, there can be at most 1 variable having $\bar{x}_a$ as a value.

{\twocolumn
\begin{figure*}

More precisely, the characteristic polynomial of the matrix is

\begin{equation}
\begin{split}
P(x)=\left( a-c-x \right)^{k-1} \left( b -e -x \right)^{p-1}\left[ x^2 - x\left( a + b + \left(k-1\right) c + \left(p -1 \right) e \right)  \right.  \\
\left. +  \left(p-1\right) ea + \left(k-1\right)cb + \left(k-1\right) \left(p-1\right)ec +ab - kpf_{1}f_{2} \right] 
\end{split}
\end{equation}

Suppose thus that the number of variables having values $\bar{x}_a$ is 1. Then, a sufficient condition for instability of the equilibrium is 

$$
\left(p-1\right)ea+ab-p f_{1} f_{2}<0
$$

Notice that in this case $f_1=f_2=e$. The sufficient condition for instability can thus be written

$$
e\left( pe - \left(p-1\right)a\right)-ab>0
$$

Replacing with the equilibrium values,

$$
\frac{-2\alpha D}{\sigma} \left( p\frac{-2\alpha D}{\sigma} - \left(p-1\right)\left( 1-\frac{2}{\sigma}\left(\bar{x}_a +\alpha D \right) \right) \right) - \left( 1 -\frac{2}{\sigma}\left(\bar{x}_a +\alpha D \right) \right) \left( 1 -\frac{2}{\sigma}\left(\bar{x}_b +\alpha D \right) \right) >0
$$

$$
\left( 1 -\frac{2}{\sigma}\left(\bar{x}_a +\alpha D \right) \right) \left( \left(p-1\right)\frac{2\alpha D}{\sigma} -1 +  \frac{2}{\sigma}\left(\bar{x}_b +\alpha D \right)\right) + p \left( \frac{2\alpha D}{\sigma}\right)^2> 0
$$

$$
\left( 1 -\frac{2}{\sigma}\left(\bar{x}_a +\alpha D \right) \right) \left( p\frac{2\alpha D}{\sigma} -1 +  \frac{2 \bar{x}_b}{\sigma}\right) + p \left( \frac{2\alpha D}{\sigma}\right)^2> 0
$$

$$
 p\frac{2\alpha D}{\sigma} \left(  1 -\frac{2 \bar{x}_a}{\sigma} \right) + \left( \frac{2 \bar{x}_b}{\sigma} -1 \right) \left( 1 -\frac{2}{\sigma}\left(\bar{x}_a +\alpha D \right) \right) > 0
$$

$$
\left(  1 -\frac{2 \bar{x}_a}{\sigma} \right) \left( p\frac{2\alpha D}{\sigma} + \frac{2 \bar{x}_b}{\sigma} -1 \right) - \frac{2\alpha D}{\sigma} \left(\frac{2 \bar{x}_b}{\sigma} -1\right)>0
$$

$$
\frac{2\alpha D}{\sigma}  \left( p+1 - \frac{2}{\sigma}\left(p \bar{x}_a + \bar{x}_b\right) \right) + \left(  1 -\frac{2 \bar{x}_a}{\sigma} \right) \left( \frac{2 \bar{x}_b}{\sigma} -1 \right) > 0
$$

\end{figure*}
}

\clearpage

The first term is positive because the values of $\bar{x}_a$ and $\bar{x}_b$ are symmetrical with respect to $\sigma/2$. The second term is also positive, and the sufficient condition for the instability of the equilibrium is thus met.

Thus, there is no stable equilibrium with non-0 variables having different values.

\section{Methods}

\subsection{Numerical integration}
All integration was performed with a custom-written implementation of the 4th-order adaptative stepsize Runge-Kutta algorithm \citep{recipes_in_c}, with $10^{-3}$ relative accuracy. Source code is available at \url{http://www-timc.imag.fr/Olivier.Cinquin/ada/ada_blas_runge_kutta.html}.
The data was plotted using GMV or gnuplot.

\subsection{Computation of convergence times}
\label{convergence_times}
A custom program was written to do the following, starting from a regular 200*200 grid of initial conditions (for 2D systems), or a 50*50*50 grid (for 3D systems), with $\forall i\ne j,\ x_i\ne x_j$, to avoid reaching unstable steady-states:
(1) integrate the system until a steady-state is reached (as defined by the sum of the absolute values of the derivative vector elements begin lower than $10^{-4}$)
(2) start the integration again, with the same initial conditions, stopping when the system gets close enough to the previous steady-state (each variable with 10\% of its steady-state value if it's not 0, lower than 0.15 if it is 0; moderate changes in these arbitrary values do not significantly affect the results). The stepsize of the Runge-Kutta algorithm was kept lower than 0.3.

\subsection{Simulations with time-dependent parameters}
In order for the system to leave steady states which had become unstable because of changed parameters, small random perturbations were applied (each variable was multiplied by a random number uniformly chosen in [0.99 .. 1.01] every 30 time units).
\end{document}